\documentclass[12pt,letter,doublespace]{article}
\usepackage{amssymb}
\usepackage{natbib,graphicx,setspace,lscape,longtable}
\usepackage{natbib,epsfig,graphicx}
\usepackage{amsmath,amsthm,amssymb,color}
\usepackage{hyperref}
\RequirePackage[mathlines, displaymath]{lineno}
\usepackage{multirow}
\bibpunct{(}{)}{;}{a}{,}{,}

\newcommand{\csection}[1]
    {\begin{center}
        \stepcounter{section}
        {\bf\large\arabic{section}. #1}
    \end{center}
    \vspace{-0.15 cm}
}

\newcommand{\scsection}[1]
    {\begin{center}
        {\bf\large #1}
    \end{center}
    \vspace{-0.15 cm}
}

\newcommand{\csubsection}[1]{
\vspace{-0.25 cm}
\begin{center}
\stepcounter{subsection}
{\it\arabic{section}.\arabic{subsection}. #1}
\end{center}
\vspace{-0.25 cm}
}

\newcommand{\scsubsection}[1]{
\vspace{-0.25 cm}
\begin{center}
\stepcounter{subsection}
{\it #1}
\end{center}
\vspace{-0.25 cm}}

\def\beqr{\begin{eqnarray}}
\def\eeqr{\end{eqnarray}}
\def\beqrs{\begin{eqnarray*}}
\def\eeqrs{\end{eqnarray*}}
\def\bep{\begin{prop}}
\def\eep{\end{prop}}
\def\bc{\begin{center}}
\def\ec{\end{center}}
\def\mR{\mathbb{R}}

\def\hDash{\bot\!\!\!\bot}

\def \dcov {\mathrm{dcov}}
\def \dcorr {\mathrm{dcorr}}

\def\mR{\mathbb{R}}

\def\calA{\mathcal{D}}
\def\calI{\mathcal{I}}

\newcommand{\bfSig}{\mathbf{\Sigma}}
\newcommand{\bb}{{\mbox{\boldmath $\beta$}}}

\newcommand{\bt}{\mathbf{t}}
\newcommand{\bs}{\mathbf{s}}
\newcommand{\bu}{\mathbf{u}}
\newcommand{\bv}{\mathbf{v}}
\newcommand{\tu}{\widetilde{\bu}}
\newcommand{\tv}{\widetilde{\bv}}
\newcommand{\x}{\mathbf{x}}

\newcommand{\y}{\mathbf{y}}

\newcommand{\pr}{\mbox{Pr}}

\newcommand{\trans}{^{\mbox{\tiny{T}}}}

\numberwithin{equation}{section}
\newtheorem{theo}{\bf Theorem}
\newtheorem{lemm}{\bf Lemma}

\renewcommand{\baselinestretch}{1.7}
\begin{document}
%\linenumbers

\title{Feature Screening via Distance Correlation Learning}
\author{\sc Runze Li, Wei Zhong and Liping Zhu\\
The Pennsylvania State University, Xiamen University \\
\& Shanghai University of Finance and Economics}
\date{\today}
\maketitle{}

\begin{singlespace}
\begin{footnotetext}[1]
{Runze Li is Professor, Department of
Statistics and The Methodology Center, The Pennsylvania State University,
University Park, PA 16802-2111. Email: \href{mailto:rli@stat.psu.edu}{rli@stat.psu.edu.}
His research was supported by National Institute on Drug Abuse (NIDA) grant
P50-DA10075 and National Natural Science Foundation of China (NNSFC) grant 11028103.
%{Wei Zhong is a graduate student,
%Department of Statistics, The Pennsylvania State University,
%University Park, PA 16802-2111. Email: wxz123@psu.edu.}
Wei Zhong is the corresponding author and Assistant Professor of Wang Yanan Institute for Studies in Economics,
Department of Statistics and Fujian Key Laboratory of Statistical Science, Xiamen University,
China. Email: \href{mailto:wxz123@psu.edu}{wxz123@psu.edu.}
His research was supported by a NIDA grant P50-DA10075 as a graduate research assistant during his graduate study,
and by the NNSFC grant 71131008 (Key Project).
Liping Zhu is Associate Professor of School of Statistics and Management,
Shanghai University of Finance and Economics, China. Email:  \href{mailto:zhu.liping@mail.shufe.edu.cn}{zhu.liping@mail.shufe.edu.cn.}
His research was supported by NNSFC grant 11071077 and a NIDA grant R21-DA024260.
%{\color{blue}The authors are grateful to Dr Yichao Wu  for sharing the ideas
%through personal communication about
%the iterative screening approach presented in this paper.}
{All authors equally contribute to this paper,
and the authors are listed in the alphabetic order. The authors thank the Editor, the AE and reviewers for their constructive comments,
which have led to a dramatic improvement of the earlier version of this paper.}
The content is solely the responsibility of the
authors and does not necessarily represent the official views of the NSF or  NIDA. }
\end{footnotetext}
\end{singlespace}

\begin{abstract}
This paper is concerned with screening features in ultrahigh dimensional data analysis, which
has become increasingly  important in diverse scientific fields.
We develop a sure independence screening procedure based on the
distance correlation (DC-SIS, for short).
%The distance correlation was proposed  by \cite{Szekely:Rizzo:Bakirov:2007}
%for measuring dependence between two random vectors.
The DC-SIS can be implemented as easily as the
sure independence screening procedure based on the Pearson correlation (SIS, for short)
proposed by \cite{Fan:Lv:2008}.
However, the DC-SIS can significantly improve the SIS. \cite{Fan:Lv:2008} established the sure screening
property for the SIS based on linear models, but the sure screening property is valid
for the DC-SIS under more general settings including linear models.
Furthermore, the implementation of the DC-SIS does not require model specification (e.g., linear model or
generalized linear model) for responses or predictors. This is a very appealing
property in ultrahigh dimensional data analysis. Moreover, the DC-SIS can be used directly to screen
grouped predictor variables and for multivariate response variables. We establish the sure screening property
for the DC-SIS, and conduct simulations to examine its finite sample performance.
Numerical comparison indicates that the DC-SIS performs much better than the SIS in various
models. We also illustrate the DC-SIS through a real data example.
\end{abstract}

\

\noindent{\bf Key words}:  Distance correlation,
%LASSO,
%ranking consistency,
%SCAD,
sure screening property, ultrahigh dimensionality, variable selection.

\

\noindent{Running Head: Distance Correlation Based SIS}

\newpage

\pagestyle{plain}
\setcounter{page}{1}
\csection{INTRODUCTION}

Various regularization methods have been proposed for feature selection in high dimensional data analysis,
which has become increasingly
frequent and important in various research fields.
These methods
include, but are not limited to, the LASSO \citep{Tibshirani:1996}, the SCAD
\citep{Fan:Li:2001,Kim:Choi:Oh:2008,Zou:Li:2008},
the LARS algorithm \citep{Efron:Hastie:Johnstone:Tibshirani:2004}, the elastic net
\citep{Zou:Hastie:2005,Zou:Zhang:2009},
the adaptive LASSO \citep{Zou:2006}  and the Dantzig selector \citep{Candes:Tao:2007}. All these methods allow
 the number of predictors to be greater than the sample size, and perform quite well for high dimensional data.

With the advent of modern technology for data collection, researchers are able to collect ultrahigh dimensional data
at relatively low cost in diverse fields of scientific research. The aforementioned
regularization methods  may not
perform well for ultrahigh dimensional data due to the simultaneous challenges of computational expediency,
statistical accuracy and algorithmic stability \citep{Fan:Samworth:Wu:2009}. These challenges call for
new statistical modeling techniques for ultrahigh dimensional data. \cite{Fan:Lv:2008} proposed the SIS
and showed that the Pearson correlation
ranking procedure possesses a sure  screening property for
linear regressions with Gaussian predictors and
responses. That is, all truly important predictors can be selected
with probability approaching one as the sample size diverges to $\infty$.
\cite{Hall:Miller:2009}  extended Pearson correlation learning by considering polynomial transformations
of predictors. To rank the importance of each predictor,
they  suggested a bootstrap procedure.
%However, how to choose an optimal transformation remains an open
%issue and is often difficult. In addition, the implementation of bootstrap is
%computationally expensive.
\cite{Fan:Samworth:Wu:2009}  and \cite{Fan:Song:2009}  proposed a more general version of independent learning
which ranks the maximum marginal likelihood estimators or the maximum
marginal likelihood  for generalized linear models.
%It contains \cite{Fan:Lv:2008} as a  special case.
\cite{Fan:Feng:Song:2011}
considered nonparametric independence screening in sparse ultrahigh
dimensional additive models. They suggested estimating the
nonparametric components marginally with spline approximation, and
ranking the importance of predictors using the magnitude of
nonparametric components. They also demonstrated that this procedure
possesses the sure screening property with vanishing false selection
rate.
%{\color{red}\cite{Wang:2009} also proposed a variable screening method, called forward regression (FR),
%to identify the relevant predictors consistently even when $p \gg n$.}
\cite{Zhu:Li:Li:Zhu:2011} proposed a sure independent ranking and screening (SIRS) procedure
to screen significant predictors in multi-index models.
%a model-free independence screening procedure, termed as
They further show that under linearity condition assumption on the predictor vector,
the SIRS enjoys the ranking consistency property (i.e, the SIRS can rank the important predictors
in the top asymptotically).
%even when the dimension $p$ diverges much faster than the sample size $n$.
% The conditions under which the independence learning
%possesses ranking consistency property are simple.
%This justifies the applicability of  independence screening procedure
%using Pearson correlation in a very wide spectrum. %The SIRS procedure
%inherits the merits of the Pearson correlation such as
%computational feasibility.
%However, none of the aforementioned independent screening
% procedures, including the model-free screening procedure SIRS,
% can handle grouped predictors or identify important predictors associated with the correlations between
%multivariate response variables.}
\cite{Ji:Jin:2010}  proposed the two-stage method: screening by Univariate thresholding
and cleaning by Penalized least squares for Selecting variables, namely UPS.  They further theoretically
demonstrated that under certain settings, the UPS can outperform the LASSO and subset selection,
 both of which are one-stage approaches. This motivates us to develop more effective screening procedures using
two-stage approaches.

In this paper, we propose a new feature screening procedure for ultrahigh dimensional data based on
distance correlation. \cite{Szekely:Rizzo:Bakirov:2007}
and \cite{Szekely:Rizzo:2009} showed that the
distance correlation of two random vectors equals to zero if and only if these two random vectors
are independent. Furthermore, the distance correlation of two univariate normal random variables
is a strictly increasing function of the absolute value of the Pearson correlation of these two normal
random variables. These two remarkable properties   motivate us to use the distance
correlation for feature screening in ultrahigh dimensional data. We refer to our Sure Independence
Screening procedure based on the Distance Correlation as the DC-SIS.
The DC-SIS can be implemented as easily as the SIS. It is   equivalent to
the SIS when both the response  and predictor variables are normally distributed.
However, the DC-SIS has appealing features that existing screening procedures including SIS do not possess.
For instance, none of the aforementioned screening procedures can handle grouped predictors or
multivariate responses. The proposed DC-SIS can be directly employed for screening grouped variables,
and it can be directly utilized for ultrahigh dimensional data with multivariate responses.
Feature screening for multivariate responses and/or grouped predictors is of great interest
in pathway analyses. As in \cite{Chenetal:2011}, pathway here means sets of proteins that are
relevant to specific biological functions without regard to the state of knowledge concerning the
interplay among such protein. Since proteins may work
interactively to perform various biological functions, pathway analyses complement the
marginal association analyses for individual protein, and aim to detect a priori defined set of proteins
that are associated with phenotypes of interest. There is a surged interest in pathway analyses
in the recent literature
\citep{Ashburneretal:2000, Moothaetal:2003,Subramanianetal:2005,Tianetal:2005, Bildetal:2006,Efron:Tibshirani:2007,Jonesetal:2008}.
Thus, it is of importance to develop feature screening procedures for multivariate
responses and/or grouped predictors.

We systematically study the theoretic properties of the DC-SIS, and prove that the DC-SIS possesses
the sure screening property in the terminology of \cite{Fan:Lv:2008} under very general model settings
including linear regression models, for which \cite{Fan:Lv:2008} established the sure screening property of the SIS.
The sure screening property is a desirable property for feature screening in ultrahigh dimensional data.
Even importantly, the DC-SIS can be used for screening features without specifying a regression model between
the response and the predictors. Compared with the model-based screening procedures
\citep{Fan:Lv:2008, Fan:Samworth:Wu:2009, Wang:2009, Fan:Song:2009,Fan:Feng:Song:2011},
the DC-SIS is a model-free screening procedure. This virtue makes the proposed
procedure robust to model mis-specification. This is a very appealing feature of the proposed procedure
in that it may be very difficult in specifying an appropriate regression model for the response and the predictors
with little information about the actual model in ultrahigh dimensional data.

We conduct Monte Carlo simulation studies to numerically compare the DC-SIS with  the SIS and SIRS.
Our simulation results indicate that the DC-SIS can significantly outperform the SIS
and the SIRS under many model settings. %See Example 1 for more details.
We also assess the performance of the DC-SIS as a grouped variable screener,
and the simulation results show that the DC-SIS performs very well.
We further examine the performance of the DC-SIS for feature screening in ultrahigh dimensional
data with multivariate responses; simulation results demonstrate that screening features for multiple
responses jointly may have dramatic advantage over screening features with each response separately.
%\cite{Fan:Lv:2008} developed an iterative SIS (ISIS) which   performs
%much better than the SIS  when  some important
%predictors are marginally independent of
%the response. The development of DC-ISIS  is indeed challenging and interesting because,
%unlike the SIS, we do not want to specify a regression model for the response and the predictors.
%Following \cite{Zhu:Li:Li:Zhu:2011}, we further propose an iterative DC-SIS procedure (DC-ISIS).
%We also conduct a Monte Carlo simulation to examine the finite sample performance of DC-ISIS and
%demonstrate that the DC-ISIS is a dramatic improvement over the DC-SIS.

The rest of this paper is organized as follows. In Section 2, we develop
the DC-SIS for feature screening and establish its   sure
screening property.  In Section 3, we examine the finite sample performance of the DC-SIS via Monte Carlo simulations.
We also illustrate the proposed methodology through a real data example.
%In Section 4, we  propose the DC-ISIS to further enhance the finite sample
%performance of the DC-SIS.
This paper concludes with a brief discussion in Section 4.
All technical proofs are given in the Appendix.

\csection{INDEPENDENCE SCREENING USING DISTANCE CORRELATION}

\csubsection{Some Preliminaries}

\cite{Szekely:Rizzo:Bakirov:2007}
advocated using the distance correlation for measuring dependence
between two random vectors. To be precise,
let $\phi_{\bu}(\bt)$ and $\phi_{\bv}(\bs)$ be the respective
 characteristic functions of  the random vectors $\bu$ and $\bv$,
 and
$\phi_{\bu,\bv}(\bt,\bs)$ be the joint characteristic function of   $\bu$ and $\bv$.
They defined the distance covariance
between $\bu$ and $\bv$ with
finite first moments to be the nonnegative number $\dcov(\bu,\bv)$ given by
\begin{eqnarray}\label{dcov1}
\dcov^2(\bu,\bv) = \int_{R^{d_u+d_v}}
\left\|\phi_{\bu,\bv}(\bt,\bs) - \phi_{\bu}(\bt)\phi_{\bv}(\bs)\right\|^2 w(\bt,\bs)
\,d\bt\, d\bs,
\end{eqnarray}
where $d_u$ and $d_v$ are the dimensions of $\bu$ and $\bv$, respectively, and
\[
w(\bt,\bs) = \left\{c_{d_u} c_{d_v} \|\bt\|_{d_u}^{1+d_u} \|\bs\|_{d_v}^{1+d_v}\right\}^{-1}
\]
with $c_d = \pi^{(1+d)/2}/\Gamma\{(1+d)/2\}$.
Throughout this paper,  $\|{\bf a}\|_d$ stands for the Euclidean norm of ${\bf a} \in \mR^d$, and
$\|\phi\|^2 = \phi\bar{\phi}$ for a complex-valued function $\phi$ with $\bar{\phi}$ being the conjugate of $\phi$.
%The integrals at 0 and $\infty$ in (\ref{dcov1}) are meant in
%the principal value sense: $\underset{\eta\to 0}{\lim}\int_{R^{d_u+d_v}\setminus\{\eta B + \eta^{-1} B^c\}}$, where $B$ is the unit
%ball in $R ^{d_u+d_v}$ with center at 0 and $B^c$ stands for the complement of $B$.
The distance correlation (DC) between $\bu$ and $\bv$ with
finite first moments  is defined as
\begin{eqnarray}\label{def7}
\dcorr(\bu,\bv) = \frac{\dcov(\bu,\bv)}{\sqrt{\dcov(\bu,\bu)\dcov(\bv,\bv)}}.
\end{eqnarray}
\cite{Szekely:Rizzo:Bakirov:2007}  systematically studied the theoretic properties of the DC.

Two remarkable properties of the DC motivate us to utilize it in a feature
screening procedure. The first one is the relationship between the DC and the
Pearson correlation coefficient. For two univariate normal random variables $U$ and $V$
with the Pearson correlation coefficient $\rho$, \cite{Szekely:Rizzo:Bakirov:2007}
{and
\cite{Szekely:Rizzo:2009}} showed that
\beqr\label{pearson}
\dcorr(U,V)  =
\left\{\frac{\rho\arcsin(\rho)+\sqrt{1-\rho^2}-\rho\arcsin(\rho/2)-\sqrt{4-\rho^2}+1}{1+\pi/3-\sqrt 3}\right\}^{1/2},
\eeqr
which is  strictly increasing in $|\rho|$. %Furthermore, $\dcorr(U,V)\le |\rho|$ (Theorem 7 of
%\cite{Szekely:Rizzo:Bakirov:2007}).
This property implies that the DC-based
feature screening procedure is   equivalent to the marginal Pearson correlation learning
for linear regression with normally distributed predictors and random error.  In such a situation,
  \cite{Fan:Lv:2008} showed that
the Pearson correlation learning
  has the sure screening property.

The second remarkable property of the DC is $\dcorr(\bu,\bv)=0$ if and only if $\bu$ and $\bv$
are independent \citep{Szekely:Rizzo:Bakirov:2007}. We note that  two univariate
random variables $U$ and $V$ are independent if and only if $U$ and $T(V)$,
a strictly monotone transformation  of $V$, are independent. This implies that
a DC-based feature screening procedure can be more effective than the marginal Pearson
correlation learning in the presence of nonlinear relationship between $U$ and $V$.
We will  demonstrate in the next section that a DC-based screening procedure is a model-free procedure
in that one does not need to specify a model structure between the predictors and the response.

\citet[Remark 3]{Szekely:Rizzo:Bakirov:2007} stated that
$$
\dcov^2(\bu,\bv)= S_{1}+S_{2}-2S_{3},
$$
where $S_j$, $j=1,$ 2 and 3, are defined below:
\begin{eqnarray}
S_{1} &=& E\left\{\|\bu-\tu\|_{d_u}\|\bv-\tv\|_{d_v}\right\},  \nonumber \\
S_{2} &=& E\left\{\|\bu-\tu\|_{d_u}\right\}E\left\{\|\bv-\tv\|_{d_v}\right\}, \label{def4}  \\
S_{3} &=& E\left\{E\left(\|\bu-\tu\|_{d_u} |\  \bu\right)E\left(\|\bv-\tv\|_{d_v} |\ \bv\right)\right\},
\nonumber
\end{eqnarray}
where $(\tu,\tv)$ is an independent copy of $(\bu,\bv)$.

Suppose that  $\left\{(\bu_i,\bv_i),i=1,\cdots,n\right\}$ is a random sample from
the population $(\bu,\bv)$.  \cite{Szekely:Rizzo:Bakirov:2007}
proposed to estimate $S_1$, $S_2$ and $S_3$   through  the usual moment estimation. To be precise,
\[
\widehat{S}_{1}=\frac 1{n^{2}}\sum_{i=1}^n\sum_{j=1}^n\|\bu_{i}-\bu_{j}\|_{d_u}\|\bv_i-\bv_j\|_{d_v},
\]
\begin{eqnarray*}
\widehat{S}_{2}&=&\frac 1{n^{2}}\sum_{i=1}^n\sum_{j=1}^n\|\bu_{i}-\bu_{j}\|_{d_u}
                  \frac 1{n^{2}}\sum_{i=1}^n\sum_{j=1}^n\|\bv_{i}-\bv_{j}\|_{d_v},\textrm{and}\\
\widehat{S}_{3}&=&\frac 1{n^{3}}\sum_{i=1}^n\sum_{j=1}^n\sum_{l=1}^n
\|\bu_{i}-\bu_{l}\|_{d_u}\|\bv_j-\bv_l\|_{d_v}.
\end{eqnarray*}
Thus, a natural estimator of $\dcov^2(\bu,\bv)$ is given by
\begin{eqnarray*}
\widehat{\dcov}^2(\bu,\bv) &=& \widehat{S}_{1}+\widehat{S}_{2}-2\widehat{S}_{3}.
\end{eqnarray*}
%\cite{Szekely:Rizzo:Bakirov:2007} established the weak convergence of this estimator.
Similarly, we can define the sample distance covariances
$\widehat{\dcov}(\bu, \bu)$ and $\widehat{\dcov}(\bv, \bv)$.
Accordingly, the sample distance correlation between $\bu$ and $\bv$ can be  defined by
\begin{eqnarray*}
\widehat{\dcorr}(\bu,\bv) =
\frac{\widehat{\dcov}(\bu,\bv)}{\sqrt{\widehat{\dcov}(\bu,\bu)\widehat{\dcov}(\bv,\bv)}}.
\end{eqnarray*}

\csubsection{An Independence Ranking and Screening Procedure}

In this section we propose an independence  screening procedure
built upon the  DC.
Let $\y = (Y_1,\cdots, Y_q)\trans$ be the response vector with support $\Psi_y$,
and $\x=(X_1,\ldots,X_p)\trans$ be the predictor vector.
We regard $q$ as a fixed number in this context. In an ultrahigh-dimensional
setting  the dimensionality $p$   greatly
exceeds the sample size $n$. It is thus natural to assume that only a small number of
predictors are relevant to $\y$.
Denote by $F(\y\mid\x)$    the conditional distribution function of
$\y$ given $\x$.
Without specifying a regression model, we   define {the index set of }
the active
and inactive predictors
by
\begin{eqnarray}\nonumber
\calA&=&\{k:  F(\y\mid\x)\  \mbox{functionally\ depends\ on}\ X_k\ \mbox{for\
some}\ \y \in\Psi_y\}, \\
\label{def0}
\calI&=&\{k:  F(\y\mid\x)\  \mbox{does not functionally depend   on}\ X_k\
\mbox{for any}\ \y  \in\Psi_y\}.
\end{eqnarray}
We further write $\x_{\calA} = \left\{X_k:k\in\calA\right\}$ and
$\x_{\calI} = \left\{X_k:k\in\calI\right\}$, and refer to $\x_\calA$
as an {\it active} predictor vector and   its complement $\x_{\calI}$ as
an {\it inactive} predictor vector. The index subset $\calA$ of all active
predictors or, equivalently, the index subset $\calI$ of all
inactive predictors, is the objective of our primary interest.
Definition (\ref{def0}) implies that $\y\hDash \x_{\calI}\mid \x_\calA$,
where  $\hDash$ denotes statistical independence.
That is, given $\x_\calA$, the remaining predictors $\x_\calI$ are independent
of $\y$. Thus the inactive predictors
$\x_\calI$ are redundant when the active predictors
$\x_\calA$ are known.

For ease of presentation, we write
\begin{eqnarray*}
\omega_{k}= \dcorr^2(X_k,\y),\quad\mbox{and}\quad
\widehat{\omega}_{k} = \widehat{\dcorr}^2(X_k,\y),
\textrm{ for $k=1,\cdots, p$ }
\end{eqnarray*}
% Denote
%\[
%\calI_0 = \left\{k: \mbox{$X_k$ and $\y$ are independent  statistically} \right\}.
%\]
%Thus, $\calI_0\subset \calI$, and   $\omega_k=0$ for $k\in \calI_0$ by the definition of the DC.
%This motivates us  to
based on a random sample $\{\x_i, \y_i\}$, $i=1,\ldots, n$.
We consider using $\omega_k$ as a marginal utility to rank the importance of $X_k$ at the population level.
We utilize the DC because it allows for  arbitrary regression relationship of $\y$ onto $\x$,
regardless of whether it is linear or nonlinear. The DC also permits univariate and multivariate
response, regardless of whether it is  continuous, discrete or
categorical. In addition, it  allows for groupwise predictors.
 Thus, this DC based screening procedure is  completely model-free.
{We select a set of important predictors with large $\hat{\omega}_k$. That is, we define
\[\widehat{\calA}^\star =\left\{k: \widehat \omega_k \ge cn^{-\kappa},
\textrm{ for } 1\le k\le p\right\},\]
where $c$ and $\kappa$ are  pre-specified threshold values which will be defined in
condition (C2) in the subsequent section.}
%{ At the sample level  we choose a submodel parallel to
%\cite{Fan:Song:2009} that
%\[\widehat{\calA}^\star =\left\{k: \widehat \omega_k \ge cn^{-\kappa},
%\textrm{ for } 1\le k\le p\right\},\]
%where $c$ and $\kappa$ are   pre-specified threshold values which will be defined in condition (C2) in the subsequent section}.

\csubsection{Theoretical Properties}
Next we study the theoretical properties of the proposed independence screening procedure built upon
the DC. {The following conditions  are imposed to facilitate the technical proofs,
although they may not be the weakest ones.

\begin{itemize}
\item [(C1)] Both $\x$ and $\y$ satisfy the sub-exponential tail probability uniformly in $p$. That is,
    there exists a positive constant $s_0$   such that for all  $0 < s \le 2s_0$,
\[
    \sup_p\max_{1\le k\le p}E\left\{\exp(s\|X_{k}\|_1^2)\right\}<\infty, \textrm{ and }
    E\{\exp(s\|\y\|_q^2)\}<\infty.
\]

\item [(C2)] The minimum distance correlation of active predictors satisfies
  $$\min_{k\in\calA}\omega_k \ge 2cn^{-\kappa}, \textrm{ for some constants $c>0$ and $0\le\kappa<1/2$}.$$
\end{itemize}
Condition (C1) follows immediately when $\x$ and $\y$
are bounded uniformly, or when they have multivariate normal distribution.
The normality assumption has been widely used in the area of ultrahigh dimensional
data analysis to facilitate the technical derivations. See, for example, \cite{Fan:Lv:2008} and \cite{Wang:2009}.
%  \cite{Bickel:Levina:2008} also assumed  the normality condition  when
%they studied the regularized estimation of large covariance matrix.

Next we explore condition (C2).
When $\x$ and $\y$ have multivariate normal distribution,
(\ref{pearson}) gives an explicit relationship between the DC
and the squared Pearson correlation. For simplicity,
we write $\dcorr(X_k,\y) = T_0\left(|\rho(X_k,\y)|\right)$
where $T_0(\cdot)$ is strictly increasing given in $(\ref{pearson})$.
In this situation, condition  (C2) requires essentially that
$\underset{k\in\calA}{\min}|\rho(X_k,\y)| \ge T_{inv}(2cn^{-\kappa})$,
where  $T_{inv}(\cdot)$  is the inverse function of $T_0(\cdot)$. This
is  parallel to   condition 3 of  \cite{Fan:Lv:2008}
where it is  assumed that
$\underset{k\in\calA}{\min}|\rho(X_k,\y)| \ge  2cn^{-\kappa}$.
This intuitive illustration implies that
condition (C2) requires that the marginal DC of active predictors
cannot be too small, which is similar to condition 3 of \cite{Fan:Lv:2008}.
 We remark here that,
although we illustrate the intuition by assuming that $\x$ and $\y$
are multivariate normal,
we do not require this assumption explicitly in our context.
The following theorem establishes the sure screening property for the DC-SIS
procedure. %which is a desired property for high-dimensional statistical learning.

{\theo \label{thm:main1} Under condition (C1), for any $0<\gamma<1/2-\kappa$,
there exist positive constants $c_1>0$ and $c_2>0$ such that
\begin{eqnarray}\label{ranking3}
\pr\left(\max_{1\le k\le p}\left|\widehat\omega_k - \omega_k\right| \ge cn^{-\kappa}\right)
\le O\left( p\left[\exp\left\{-c_1 n^{1-2(\kappa+\gamma)} \right\}+n\exp\left(-c_2 n^{\gamma}\right)\right] \right).
\end{eqnarray}
Under conditions (C1) and (C2), we have that
\begin{eqnarray}\label{ranking4}
\pr\left(\calA \subseteq\widehat{\calA}^\star\right)\ge
1 - O\left(s_n\left[\exp\left\{-c_1 n^{1-2(\kappa+\gamma)} \right\}+n\exp\left(-c_2 n^{\gamma}\right)\right]\right),
\end{eqnarray}
where $s_n$ is the cardinality of $\calA$.}

The sure screening property holds for the DC-SIS
 under milder conditions than those for the SIS \citep{Fan:Lv:2008} in that
we do not require the regression function of $\y$ onto $\x$ to be linear.
Thus, the DC-SIS provides a unified alternative to existing model-based sure screening procedures.
Compared with the SIRS,
%We recall that both DC-SIS and SIRS are model-free screening procedures, and \cite{Zhu:Li:Li:Zhu:2011} established
% the ranking consistency property for the SIRS. The ranking consistency property
%  implies the sure screening property immediately. However, the DC-SIS inherits the merits of distance correlation and it has several  advantages
% over the SIS and SIRS which were built upon Pearson correlation.
%Specifically,
the DC-SIS can effectively handle grouped predictors and multivariate responses.

To balance the two terms in the right hand side of (\ref{ranking3}), we
choose the optimal order $\gamma=(1-2\kappa)/3$, then
the first part of Theorem \ref{thm:main1} becomes
$$\pr\Big(\max_{1\le k\le p}\left|\widehat\omega_k - \omega_k\right| \ge cn^{-\kappa}\Big)
\leq  O\left( p\left[\exp\left\{-c_1 n^{(1-2\kappa)/3} \right\}\right] \right), $$
for some constant $c_1>0$, indicating that we can handle the NP-dimensionality of  order
$\log{p} = o\left( n^{(1-2\kappa)/3} \right).$
If we further assume that
 $X_k$ and $\y$ are bounded uniformly in $p$, then we can obtain without much difficulty
that
$$\pr\left(\max_{1\le k\le p}\left|\widehat\omega_k - \omega_k\right| \ge cn^{-\kappa}\right)
\leq  O\left( p\left[\exp\left\{-c_1 n^{1-2\kappa} \right\}\right] \right).$$
In this case, we can handle the NP-dimensionality
$\log{p} = o\left( n^{1-2\kappa} \right).$
%That is actually the largest possible $p$ one can handle in
%theory.} ({\bf I don't understand this. Is that true?})

\

\csection{NUMERICAL STUDIES}

In this section we  assess the performance of the  DC-SIS
by Monte Carlo simulation. Our simulation studies were conducted using R code.
We further illustrate the proposed screening procedure with an empirical analysis of a real data example.

In  Examples 1, 2 and 3, we generate   $\x = (X_1,X_2,\cdots,X_p)\trans$
from  normal distribution with  zero mean and  covariance matrix
$\bfSig = (\sigma_{ij})_{p\times p}$, and the error term
$\varepsilon$ from standard normal distribution ${\cal N}(0,1)$.
{We consider two
covariance matrices to assess the performance of the DC-SIS and to compare with existing
methods: (i) $\sigma_{ij}=0.8^{|i-j|}$ and (ii) $\sigma_{ij}=0.5^{|i-j|}$.
  We fix the sample size $n$ to be 200 and vary the dimension $p$ from 2,000 to 5,000.
We repeat each experiment 500 times, and evaluate the performance through the following three criteria.}
\begin{enumerate}
    \item ${\cal S}$:  the minimum model size to include %ensure the inclusion of
     all active predictors.
          We report the 5\%, 25\%, 50\%, 75\% and 95\% quantiles of ${\cal S}$ out of 500 replications.
%          to illustrate the finite sample performance.
    \item ${\cal P}_{s}$: the proportion that %every single
     an individual active predictor is selected  for a given model size $d$ in the 500 replications.
    \item ${\cal P}_{a}$: the proportion that all active predictors are selected for
    a given model size $d$ in the 500 replications.
\end{enumerate}
The $\cal S$ is used to measure the model complexity of the resulting model of an underlying screening
procedure. The closer to the minimum model size the $\cal S$ is, the better the screening procedure is.
%{\color{red}The $\cal S$ is expected to be  close to the number of truly active predictors.}
%      In an ideal situation  where the active predictors are ranked in the top,
% $\cal S$ offers an approximation of   the cardinality of
%     $\calA$.}
     The sure screening property ensures  that ${\cal P}_s$ and ${\cal P}_{a}$ are
      both close to one when the estimated model size $d$ is sufficiently large.
   We choose  $d$ to be $d_1  =[n/\log n]$, $d_2 = 2[n/\log n]$  and $d_3 = 3[n/\log n]$ throughout
    our simulations to empirically examine the effect of the cutoff,
    where $[a]$ denotes the integer part of $a$.

\begin{table}[!h]
\begin{center}
%\caption{\label{example1.size}The minimum model size $\cal S$ in Example 1. The quintuplet
%in each table cell consists of the 5\%, 25\%, 50\%, 75\% and 95\% quantiles of ${\cal S}$ out of 500 replications.}
\caption{\label{example1.size} \small{ The 5\%, 25\%, 50\%, 75\% and 95\% quantiles of the minimum model size $\cal S$ out of 500 replications in Example 1.}}
\scalebox{0.7}{%
\begin{tabular}{r|rrrrr|rrrrr|rrrrr}
  \hline
$\cal S$  &\multicolumn{5}{c|}{SIS} & \multicolumn{5}{c|}{SIRS} & \multicolumn{5}{c}{DC-SIS} \\ \hline
Model & 5\% & 25\% & 50\% & 75\% & 95\%   & 5\% & 25\% & 50\% & 75\% & 95\%   & 5\% & 25\% & 50\% & 75\% & 95\%   \\ \hline
\multicolumn{16}{l}{{\bf case 1:} $p=2000$ and $\sigma_{ij}=0.5^{|i-j|}$}\\  %%%%%
  \hline
  \textbf{(1.a)} & 4.0 & 4.0 & 5.0 & 7.0 & 21.2 & 4.0 & 4.0 & 5.0 & 7.0 & 45.1 & 4.0 & 4.0 & 4.0 & 6.0 & 18.0 \\
  \textbf{(1.b)} & 68.0 & 578.5 & 1180.5 & 1634.5 & 1938.0 & 232.9 & 871.5 & 1386.0 & 1725.2 & 1942.4 & 5.0 & 9.0 & 24.5 & 73.0 & 345.1 \\
  \textbf{(1.c)} & 395.9 & 1037.2 & 1438.0 & 1745.0 & 1945.1 & 238.5 & 805.0 & 1320.0 & 1697.0 & 1946.0 & 6.0 & 10.0 & 22.0 & 59.0 & 324.1 \\ \textbf{(1.d)} & 130.5 & 611.2 & 1166.0 & 1637.0 & 1936.5 & 42.0 & 304.2 & 797.0 & 1432.2 & 1846.1 & 4.0 & 5.0 & 9.0 & 41.0 & 336.2 \\
   \hline
\multicolumn{16}{l}{{\bf case 2:} $p=2000$ and $\sigma_{ij}=0.8^{|i-j|}$}\\  %%%
  \hline
\textbf{(1.a)} & 5.0 & 9.0 & 16.0 & 97.0 & 729.4 & 5.0 & 9.0 & 18.0 & 112.8 & 957.1 & 4.0 & 7.0 & 11.0 & 31.2 & 507.2 \\
\textbf{(1.b)} & 26.0 & 283.2 & 852.0 & 1541.2 & 1919.0 & 103.9 & 603.0 & 1174.0 & 1699.2 & 1968.0 & 5.0 & 8.0 & 11.0 & 17.0 & 98.0 \\
\textbf{(1.c)} & 224.5 & 775.2 & 1249.5 & 1670.0 & 1951.1 & 118.6 & 573.2 & 1201.5 & 1685.2 & 1955.0 & 7.0 & 10.0 & 15.0 & 38.0 & 198.3 \\
\textbf{(1.d)} & 79.0 & 583.8 & 1107.5 & 1626.2 & 1930.0 & 50.9 & 300.5 & 728.0 & 1368.2 & 1900.1 & 4.0 & 7.0 & 17.0 & 73.2 & 653.1 \\
   \hline
\multicolumn{16}{l}{{\bf case 3:} $p=5000$ and $\sigma_{ij}=0.5^{|i-j|}$}\\    %%%%%
  \hline
\textbf{(1.a)} & 4.0 & 4.0 & 5.0 & 6.0 & 59.0 & 4.0 & 4.0 & 5.0 & 7.0 & 88.4 & 4.0 & 4.0 & 4.0 & 6.0 & 34.1 \\
\textbf{(1.b)} & 165.1 & 1112.5 & 2729.0 & 3997.2 & 4851.5 & 560.8 & 1913.0 & 3249.0 & 4329.0 & 4869.1 & 5.0 & 11.8 & 45.0 & 168.8 & 956.7 \\
\textbf{(1.c)} & 1183.7 & 2712.0 & 3604.5 & 4380.2 & 4885.0 & 440.4 & 1949.0 & 3205.5 & 4242.8 & 4883.1 & 7.0 & 17.0 & 53.0 & 179.5 & 732.0 \\
\textbf{(1.d)} & 259.9 & 1338.5 & 2808.5 & 3990.8 & 4764.9 & 118.7 & 823.2 & 1833.5 & 3314.5 & 4706.1 & 4.0 & 5.0 & 15.0 & 77.2 & 848.2 \\
   \hline
\multicolumn{16}{l}{{\bf case 4:} $p=5000$ and $\sigma_{ij}=0.8^{|i-j|}$}\\    %%%%%
  \hline
\textbf{(1.a)} & 5.0 & 10.0 & 26.5 & 251.5 & 2522.7 & 5.0 & 10.0 & 28.0 & 324.8 & 3246.4 & 5.0 & 8.0 & 14.0 & 69.0 & 1455.1 \\
\textbf{(1.b)} & 40.7 & 639.8 & 2072.0 & 3803.8 & 4801.7 & 215.7 & 1677.8 & 3010.0 & 4352.2 & 4934.1 & 5.0 & 8.0 & 11.0 & 21.0 & 162.0 \\
\textbf{(1.c)} & 479.2 & 1884.8 & 3347.5 & 4298.5 & 4875.2 & 297.7 & 1359.2 & 2738.5 & 4072.5 & 4877.6 & 8.0 & 12.0 & 22.0 & 83.0 & 657.9 \\
\textbf{(1.d)} & 307.0 & 1544.0 & 2832.5 & 4026.2 & 4785.2 & 148.2 & 672.0 & 1874.0 & 3330.0 & 4665.2 & 4.0 & 7.0 & 21.0 & 165.2 & 1330.0 \\
   \hline
\end{tabular}}
\end{center}
\end{table}

\noindent{\bf Example 1}. This example is designed to compare the finite sample performance of the DC-SIS with the SIS \citep{Fan:Lv:2008} and
SIRS \citep{Zhu:Li:Li:Zhu:2011}. In this example, we generate the response from the following four models:
  \begin{eqnarray*}
    \textbf{(1.a):\hspace{1cm}} Y&=&c_1\beta_1X_1+c_2\beta_2X_2+c_3\beta_3{\bf 1}(X_{12}<0)+c_4\beta_4X_{22}+\varepsilon,\\
    \textbf{(1.b):\hspace{1cm}} Y&=&c_1\beta_1X_1X_2+c_3\beta_2{\bf 1}(X_{12}<0)+c_4\beta_3X_{22}+\varepsilon,\\
    \textbf{(1.c):\hspace{1cm}} Y&=&c_1\beta_1X_1X_2+c_3\beta_2{\bf 1}(X_{12}<0)X_{22}+\varepsilon,\\
    \textbf{(1.d):\hspace{1cm}} Y&=&c_1\beta_1X_1+c_2\beta_2X_2+c_3\beta_3{\bf 1}(X_{12}<0)+\exp(c_4|X_{22}|)\varepsilon,
  \end{eqnarray*}
  where ${\bf 1}(X_{12}<0)$ is an indicator function.
 %  taking value $1$ when $X_{12}$ is negative and 0 otherwise.
The regression functions $E(Y\mid\x)$ in models {\bf(1.a)}-{\bf(1.d)} are all nonlinear in $X_{12}$. In addition,
models ${\bf(1.b)}$ and ${\bf(1.c)}$  contain an interaction term $X_1X_2$,
and  model ${\bf(1.d)}$ is heteroscedastic. Following Fan and Lv (2008),
we choose  $\beta_j=(-1)^{U}(a+|Z|)$ for $j=1,2,3$ and 4, where $a=4\log n/\sqrt{n}$,
$U\sim \mbox{Bernoulli}(0.4)$ and $Z\sim {\cal N}(0,1)$. {We set
$(c_1,c_2,c_3,c_4)=(2,0.5,3,2)$ in this example to challenge the feature
screening procedures under consideration.}
For each independence screening procedure, we compute the associated marginal utility between each predictor
$X_k$ and the response $Y$. That is, we regard $\x = (X_1,\ldots,X_p)\trans\in\mR^p$
as the predictor vector in this example.
 %REMARK: in the model ${\bf(1.b)}$, we compute the marginal utility between $X_1$ and $Y$, $X_2$ and $Y$,
 %separately, not the marginal utility between $X_1X_2$ and $Y$. Note that it is extremely difficult to know how the predictor $X_j$
 %is correlated to the response $Y$ in the ultrahigh dimensional framework. For the model ${\bf(1.b)}$, we can not know whether there
 %exists the interaction effect $X_1X_2$ in the initial screening stage. This is also one of main reasons that we need a model-free
 %independence screening procedure.}

Tables~\ref{example1.size} and \ref{example1.prob1} depict the simulation results for $\cal S$, ${\cal P}_{s}$ and ${\cal P}_{a}$.
The performances of the DC-SIS,  SIS and  SIRS are quite similar in model {\bf (1.a)}, indicating
that the SIS has a robust performance if the working linear model does not deviate far from the underlying true model.
The DC-SIS outperforms the SIS and SIRS significantly in models {\bf (1.b)}, {\bf (1.c)} and {\bf (1.d)}.
Both the SIS and SIRS have little chance to identify the important predictors $X_1$ and $X_2$ in models
{\bf (1.b)} and {\bf (1.c)}, and $X_{22}$ in model {\bf (1.d)}.

\begin{table}[htbp]
\begin{center}
\caption{\label{example1.prob1}  \small{The proportions of  ${\cal P}_{s}$ and  ${\cal P}_{a}$ in Example 1. The user-specified model sizes
$d_1 = [n/\log n]$, $d_2 = 2[n/\log n]$ and $d_3 = 3[n/\log n]$.}}
%The empirical probabilities of each active predictor (denoted by ${\cal P}_{s}$) and all active predictors (denoted by ${\cal P}_{a}$)
%are chosen for a given model size $d_i$, where $d_1 = [n/\log n]$, $d_2  = 2[n/\log n]$ and $d_3 = 3[n/\log n]$.   }  %%%%%
\scalebox{0.75}{%
\begin{tabular}{rc|cccc|c|cccc|c|cccc|c}
  \hline
&  &\multicolumn{5}{c|}{SIS} & \multicolumn{5}{c|}{SIRS} & \multicolumn{5}{c}{DC-SIS} \\ \hline
      &        & \multicolumn{4}{c|}{${\cal P}_{s}$} & ${\cal P}_{a}$ & \multicolumn{4}{c|}{${\cal P}_{s}$} & ${\cal P}_{a}$ & \multicolumn{4}{c|}{${\cal P}_{s}$} & ${\cal P}_{a}$ \\ \hline
model & size   & $X_1$ & $X_2$ & $X_{12}$ &  $X_{22}$ & ALL   & $X_1$ & $X_2$ & $X_{12}$ &  $X_{22}$ & ALL
               & $X_1$ & $X_2$ & $X_{12}$ &  $X_{22}$ & ALL \\ \hline
\multicolumn{17}{l}{{\bf case 1:} $p=2000$ and $\sigma_{ij}=0.5^{|i-j|}$}\\\hline  %%%%%
               & $d_1$  & 1.00 & 1.00 & 0.96 & 1.00 & 0.96 & 1.00 & 1.00 & 0.95 & 1.00 & 0.94 & 1.00 & 1.00 & 0.97 & 1.00 & 0.96 \\
\textbf{(1.a)} & $d_2$  & 1.00 & 1.00 & 0.98 & 1.00 & 0.97 & 1.00 & 1.00 & 0.96 & 1.00 & 0.96 & 1.00 & 1.00 & 0.98 & 1.00 & 0.98 \\
               & $d_3$  & 1.00 & 1.00 & 0.98 & 1.00 & 0.98 & 1.00 & 1.00 & 0.97 & 1.00 & 0.97 & 1.00 & 1.00 & 0.99 & 1.00 & 0.98 \\     \hline
               & $d_1$  & 0.08 & 0.07 & 0.97 & 1.00 & 0.03 & 0.02 & 0.03 & 0.98 & 1.00 & 0.00 & 0.72 & 0.70 & 0.99 & 1.00 & 0.58 \\
\textbf{(1.b)} & $d_2$  & 0.12 & 0.13 & 0.98 & 1.00 & 0.06 & 0.05 & 0.05 & 0.99 & 1.00 & 0.01 & 0.85 & 0.84 & 1.00 & 1.00 & 0.76 \\
               & $d_3$  & 0.15 & 0.17 & 0.99 & 1.00 & 0.07 & 0.06 & 0.06 & 0.99 & 1.00 & 0.01 & 0.89 & 0.88 & 1.00 & 1.00 & 0.82 \\      \hline
               & $d_1$  & 0.12 & 0.13 & 0.01 & 0.99 & 0.00 & 0.04 & 0.03 & 0.51 & 1.00 & 0.01 & 0.93 & 0.93 & 0.77 & 1.00 & 0.65 \\
\textbf{(1.c)} & $d_2$  & 0.17 & 0.18 & 0.03 & 0.99 & 0.00 & 0.07 & 0.05 & 0.67 & 1.00 & 0.01 & 0.97 & 0.96 & 0.84 & 1.00 & 0.79 \\
               & $d_3$  & 0.21 & 0.21 & 0.05 & 0.99 & 0.00 & 0.09 & 0.08 & 0.75 & 1.00 & 0.02 & 0.98 & 0.97 & 0.89 & 1.00 & 0.84 \\      \hline
               & $d_1$  & 0.42 & 0.22 & 0.14 & 0.42 & 0.02 & 1.00 & 0.98 & 0.87 & 0.05 & 0.04 & 1.00 & 0.91 & 0.81 & 0.99 & 0.73 \\
\textbf{(1.d)} & $d_2$  & 0.48 & 0.29 & 0.22 & 0.50 & 0.03 & 1.00 & 0.99 & 0.91 & 0.10 & 0.09 & 1.00 & 0.94 & 0.87 & 1.00 & 0.82 \\
               & $d_3$  & 0.56 & 0.32 & 0.26 & 0.54 & 0.04 & 1.00 & 0.99 & 0.93 & 0.12 & 0.11 & 1.00 & 0.96 & 0.92 & 1.00 & 0.88 \\
           \hline
\multicolumn{17}{l}{{\bf case 2:} $p=2000$ and $\sigma_{ij}=0.8^{|i-j|}$}\\\hline  %%%%%

               & $d_1$  & 1.00 & 1.00 & 0.63 & 1.00 & 0.63 & 1.00 & 1.00 & 0.62 & 1.00 & 0.62 & 1.00 & 1.00 & 0.78 & 1.00 & 0.77 \\
\textbf{(1.a)} & $d_2$  & 1.00 & 1.00 & 0.71 & 1.00 & 0.72 & 1.00 & 1.00 & 0.70 & 1.00 & 0.69 & 1.00 & 1.00 & 0.84 & 1.00 & 0.84 \\
               & $d_3$  & 1.00 & 1.00 & 0.77 & 1.00 & 0.78 & 1.00 & 1.00 & 0.75 & 1.00 & 0.75 & 1.00 & 1.00 & 0.86 & 1.00 & 0.86 \\     \hline
               & $d_1$  & 0.12 & 0.13 & 0.81 & 1.00 & 0.06 & 0.04 & 0.04 & 0.88 & 1.00 & 0.02 & 0.97 & 0.98 & 0.92 & 1.00 & 0.88 \\
\textbf{(1.b)} & $d_2$  & 0.19 & 0.19 & 0.86 & 1.00 & 0.12 & 0.07 & 0.07 & 0.91 & 1.00 & 0.03 & 0.99 & 0.99 & 0.95 & 1.00 & 0.94 \\
               & $d_3$  & 0.22 & 0.23 & 0.88 & 1.00 & 0.15 & 0.09 & 0.11 & 0.93 & 1.00 & 0.06 & 1.00 & 0.99 & 0.96 & 1.00 & 0.96 \\      \hline
               & $d_1$  & 0.17 & 0.16 & 0.03 & 0.99 & 0.00 & 0.04 & 0.04 & 0.53 & 1.00 & 0.02 & 1.00 & 1.00 & 0.75 & 1.00 & 0.75 \\
\textbf{(1.c)} & $d_2$  & 0.22 & 0.22 & 0.06 & 1.00 & 0.01 & 0.08 & 0.08 & 0.71 & 1.00 & 0.03 & 1.00 & 1.00 & 0.85 & 1.00 & 0.86 \\
               & $d_3$  & 0.27 & 0.27 & 0.10 & 1.00 & 0.03 & 0.10 & 0.10 & 0.81 & 1.00 & 0.05 & 1.00 & 1.00 & 0.90 & 1.00 & 0.90 \\      \hline
               & $d_1$   & 0.44 & 0.38 & 0.11 & 0.45 & 0.03 & 1.00 & 1.00 & 0.73 & 0.05 & 0.04 & 0.99 & 0.98 & 0.68 & 1.00 & 0.67 \\
\textbf{(1.d)} & $d_2$   & 0.51 & 0.46 & 0.18 & 0.53 & 0.05 & 1.00 & 1.00 & 0.81 & 0.09 & 0.08 & 1.00 & 0.98 & 0.76 & 1.00 & 0.75 \\
               & $d_3$   & 0.55 & 0.49 & 0.22 & 0.57 & 0.06 & 1.00 & 1.00 & 0.84 & 0.14 & 0.11 & 1.00 & 0.99 & 0.80 & 1.00 & 0.80 \\
\hline
\multicolumn{17}{l}{{\bf case 3:} $p=5000$ and $\sigma_{ij}=0.5^{|i-j|}$}\\\hline  %%%%%
               & $d_1$  & 1.00 & 1.00 & 0.94 & 1.00 & 0.94 & 1.00 & 0.99 & 0.92 & 1.00 & 0.92 & 1.00 & 0.99 & 0.96 & 1.00 & 0.95 \\
\textbf{(1.a)} & $d_2$  & 1.00 & 1.00 & 0.95 & 1.00 & 0.95 & 1.00 & 1.00 & 0.95 & 1.00 & 0.95 & 1.00 & 1.00 & 0.97 & 1.00 & 0.97 \\
               & $d_3$  & 1.00 & 1.00 & 0.96 & 1.00 & 0.96 & 1.00 & 1.00 & 0.96 & 1.00 & 0.96 & 1.00 & 1.00 & 0.98 & 1.00 & 0.98 \\      \hline    & $d_1$  & 0.06 & 0.06 & 0.94 & 1.00 & 0.02 & 0.02 & 0.02 & 0.96 & 1.00 & 0.00 & 0.59 & 0.60 & 0.98 & 1.00 & 0.46 \\
\textbf{(1.b)} & $d_2$  & 0.09 & 0.09 & 0.96 & 1.00 & 0.03 & 0.03 & 0.03 & 0.97 & 1.00 & 0.01 & 0.72 & 0.72 & 0.99 & 1.00 & 0.61 \\
               & $d_3$  & 0.12 & 0.10 & 0.97 & 1.00 & 0.04 & 0.05 & 0.04 & 0.98 & 1.00 & 0.01 & 0.79 & 0.78 & 0.99 & 1.00 & 0.68 \\       \hline
               & $d_1$  & 0.06 & 0.06 & 0.01 & 0.99 & 0.00 & 0.03 & 0.02 & 0.30 & 1.00 & 0.00 & 0.86 & 0.87 & 0.61 & 1.00 & 0.41 \\                \textbf{(1.c)} & $d_2$  & 0.10 & 0.10 & 0.02 & 1.00 & 0.00 & 0.04 & 0.03 & 0.45 & 1.00 & 0.00 & 0.92 & 0.93 & 0.69 & 1.00 & 0.57 \\
               & $d_3$  & 0.12 & 0.12 & 0.02 & 1.00 & 0.00 & 0.05 & 0.05 & 0.53 & 1.00 & 0.00 & 0.94 & 0.95 & 0.73 & 1.00 & 0.64 \\       \hline
               & $d_1$   & 0.39 & 0.21 & 0.11 & 0.40 & 0.01 & 1.00 & 0.97 & 0.82 & 0.02 & 0.02 & 0.99 & 0.87 & 0.74 & 0.99 & 0.65 \\
\textbf{(1.d)} & $d_2$   & 0.44 & 0.24 & 0.14 & 0.45 & 0.01 & 1.00 & 0.98 & 0.88 & 0.04 & 0.03 & 0.99 & 0.90 & 0.81 & 0.99 & 0.75 \\               & $d_3$   & 0.48 & 0.28 & 0.17 & 0.47 & 0.02 & 1.00 & 0.99 & 0.90 & 0.06 & 0.05 & 0.99 & 0.92 & 0.85 & 1.00 & 0.79 \\
   \hline
\multicolumn{17}{l}{{\bf case 4:} $p=5000$ and $\sigma_{ij}=0.8^{|i-j|}$}\\\hline  %%%%%
               & $d_1$  & 1.00 & 1.00 & 0.55 & 1.00 & 0.55 & 1.00 & 1.00 & 0.55 & 1.00 & 0.55 & 1.00 & 1.00 & 0.70 & 1.00 & 0.69 \\
\textbf{(1.a)} & $d_2$  & 1.00 & 1.00 & 0.61 & 1.00 & 0.62 & 1.00 & 1.00 & 0.61 & 1.00 & 0.61 & 1.00 & 1.00 & 0.76 & 1.00 & 0.76 \\
               & $d_3$  & 1.00 & 1.00 & 0.67 & 1.00 & 0.67 & 1.00 & 1.00 & 0.64 & 1.00 & 0.64 & 1.00 & 1.00 & 0.80 & 1.00 & 0.80 \\         \hline      & $d_1$  & 0.10 & 0.09 & 0.74 & 1.00 & 0.05 & 0.02 & 0.02 & 0.83 & 1.00 & 0.00 & 0.94 & 0.94 & 0.90 & 1.00 & 0.82 \\
\textbf{(1.b)} & $d_2$  & 0.12 & 0.13 & 0.81 & 1.00 & 0.07 & 0.03 & 0.04 & 0.87 & 1.00 & 0.01 & 0.97 & 0.97 & 0.93 & 1.00 & 0.89 \\
               & $d_3$  & 0.15 & 0.16 & 0.84 & 1.00 & 0.10 & 0.05 & 0.06 & 0.90 & 1.00 & 0.02 & 0.98 & 0.98 & 0.95 & 1.00 & 0.92 \\          \hline
               & $d_1$  & 0.10 & 0.10 & 0.02 & 0.98 & 0.00 & 0.02 & 0.03 & 0.34 & 1.00 & 0.00 & 1.00 & 1.00 & 0.64 & 1.00 & 0.63 \\                     \textbf{(1.c)} & $d_2$  & 0.13 & 0.14 & 0.04 & 0.99 & 0.01 & 0.04 & 0.04 & 0.50 & 1.00 & 0.01 & 1.00 & 1.00 & 0.74 & 1.00 & 0.74 \\
               & $d_3$  & 0.16 & 0.18 & 0.05 & 0.99 & 0.01 & 0.05 & 0.05 & 0.61 & 1.00 & 0.02 & 1.00 & 1.00 & 0.79 & 1.00 & 0.79 \\          \hline
               & $d_1$   & 0.42 & 0.32 & 0.09 & 0.40 & 0.01 & 1.00 & 1.00 & 0.66 & 0.02 & 0.01 & 0.99 & 0.97 & 0.63 & 0.98 & 0.59 \\
\textbf{(1.d)} & $d_2$   & 0.48 & 0.39 & 0.12 & 0.44 & 0.02 & 1.00 & 1.00 & 0.74 & 0.04 & 0.03 & 0.99 & 0.97 & 0.70 & 1.00 & 0.68 \\                    & $d_3$   & 0.51 & 0.42 & 0.15 & 0.46 & 0.02 & 1.00 & 1.00 & 0.78 & 0.05 & 0.04 & 0.99 & 0.98 & 0.73 & 1.00 & 0.71 \\
   \hline
\end{tabular}}
\end{center}
\end{table}

%%%%%%%%%%%%   Example 2  %%%%%%%%%%%%%%%%%%%%%%
\noindent{\bf Example 2.}
We illustrate that the DC-SIS can be directly used for screening grouped predictors.
%Next we apply the DC-SIS to regressions with group-variables.
In many regression problems, some predictors can be naturally grouped.
The most common example which contains group variables
is the multi-factor ANOVA problem, in which each factor may have several levels
and can be expressed through a group of dummy variables. The goal of ANOVA is  to select
important main effects and interactions for accurate predictions, which amounts to the selection
of groups of  dummy variables.
To demonstrate the   practicability of the DC-SIS, we adopt
the following model:
 \begin{eqnarray*}
    Y &=& c_1\beta_1X_1+c_2\beta_2X_2
    +c_3\beta_3 \{{\bf1}(X_{12}<q_1)
    +1.5\times{\bf1}(q_1 \leq X_{12}<q_2)
    \\&&+2\times{\bf1}(q_2 \leq X_{12}<q_3) \}
    +c_4\beta_4X_{22}+\varepsilon,
  \end{eqnarray*}
  where   $q_1$, $q_2$  and $q_3$ are the  25\%, 50\% and 75\% quantiles of $X_{12}$, respectively.
The variables $X$  with the coefficients $c_i$'s and $\beta_i$'s are the same as those in Example 1.
We write
 \[\widetilde\x_{12}=\left\{{\bf1}(X_{12}<q_1),\ {\bf1}(q_1 \leq X_{12}<q_2), \
{\bf1}(q_2 \leq X_{12}<q_3))\right\}\trans.\]
These three correlated variables  naturally become a group. The predictor vector
in this example becomes
$\x = (X_1,\ldots,X_{11},\widetilde\x_{12},X_{13},\ldots, X_p)\trans{\in\mR^{p+2}}$.
We remark here that the marginal utility of the grouped variable
 $\widetilde\x_{12}$ is defined  by
\[
\widehat{\omega}_{12} = \widehat{\dcorr}^2(\widetilde\x_{12},Y).
\]
The 5\%, 25\%, 50\%, 75\% and 95\% percentiles of the minimum model size ${\cal S}$ are
{summarized in  Table \ref{example2.size}}. These percentiles indicate  that with very high probability,
 the minimum model size $\cal S$ to ensure the inclusion of all active
predictors is small. Note that $[n/\log(n)]=37$.
Thus, almost all ${\cal P}_{s}$s and ${\cal P}_{a}$s equal 100\%.
All active predictors including the grouped variable  $\widetilde\x_{12}$  can 
almost perfectly be selected into the resulting model across all three different 
model sizes. Hence, the DC-SIS is  efficient to select the grouped predictors.

\begin{table}[!h]
\begin{center}
%\caption{\label{example2.size}\small{The minimum model size $\cal S$ in Example 2. The quintuplet
%in each table cell consists of the 5\%, 25\%, 50\%, 75\% and 95\% quantiles of ${\cal S}$ out of 500 replications.}}
\caption{\label{example2.size}\small{ The 5\%, 25\%, 50\%, 75\% and 95\% quantiles of the minimum model size $\cal S$ out of 500 replications in Example 2.}}
\scalebox{0.8}{%
\begin{tabular}{r|rrrrr|rrrrr}
  \hline
$\cal S$  &\multicolumn{5}{c|}{$p=2000$} & \multicolumn{5}{c}{$p=5000$}  \\    \hline
  & 5\% & 25\% & 50\% & 75\% & 95\% & 5\% & 25\% & 50\% & 75\% & 95\% \\
\hline
$\sigma_{ij}=0.5^{|i-j|}$ & 4.0 & 4.0 & 4.0 & 5.0 & 12.0  & 4.0 & 4.0 & 4.0 & 6.0 & 16.1   \\
$\sigma_{ij}=0.8^{|i-j|}$ & 4.0 & 5.0 & 7.0 & 9.0 & 15.2  & 4.0 & 5.0 & 7.0 & 9.0 & 21.0  \\
   \hline
\end{tabular}}
\end{center}
\end{table}

\noindent{\textbf{Example 3.}}  In this example, we investigate
the performance of the DC-SIS with multivariate responses.
The SIS proposed in \cite{Fan:Lv:2008} cannot be directly applied for such settings.
%In such a situation  it is  difficult to screen important variables
%with the current well-established screening methods such as SIS by \cite{Fan:Lv:2008}.
In contrast, the DC-SIS is ready for  screening the active predictors by the nature of DC.
 In this example, we generate  $\y=(Y_1,Y_2)\trans$ from normal distribution with mean zero and
covariance matrix $\bfSig_{\y\mid\x}=(\sigma_{\x,ij})_{2\times 2}$, where
$\sigma_{\x,11} = \sigma_{\x,22} =1$ and
$\sigma_{\x,12} = \sigma_{\x,21} =\sigma(\x)$.
%\begin{eqnarray*}
%\left(
%  \begin{array}{c}
%    Y_1 \\
%    Y_2 \\
%  \end{array}
%\right)\sim {\cal N}\left(
%\left(
%  \begin{array}{c}
%    0 \\
%    0 \\
%  \end{array}
%\right),
%\left(
%  \begin{array}{cc}
%    1 & \sigma^2(\bb\trans\x) \\
%    \sigma^2(\bb\trans\x) & 1 \\
%  \end{array}
%\right)\right)
%\end{eqnarray*}
We consider two scenarios for the correlation function $\sigma(\x)$:
\begin{description}
\item {\bf(3.a):}   $\sigma(\x)=\sin(\bb_1\trans \x)$, where $\bb_1=(0.8,0.6,0,\ldots,0)\trans$.
\item {\bf(3.b):}   $\sigma(\x)=\left\{\exp(\bb_2\trans \x)-1\right\}/\left\{\exp(\bb_2\trans \x)+1\right\}$,
where $\bb_2=(2-U_1,2-U_2,2-U_3,2-U_4,0,\ldots,0)\trans$ with
$U_i$'s being independent and identically distributed according to uniform distribution $\textrm{Uniform}[0,1]$.
\end{description}
%Similar to our previous examples,
%we generate  $U_i$'s  independently from uniform distribution $\textrm{Uniform}[0,1]$.
Tables \ref{example3.size} and \ref{example3.prob} depict the simulation results.
Table~\ref{example3.size} implies that the DC-SIS performs reasonably well for both models ({\bf 3.a}) and ({\bf 3.b}) in
terms of model complexity.
Table \ref{example3.prob} indicates that the   proportions
that the active predictors are selected into the model are close to one,
  which supports the assertion that the DC-SIS processes the sure screening property.
It implies  that the DC-SIS can identify the active predictors contained in correlations between multivariate responses.
This may be potentially useful in gene co-expression analysis.

\begin{table}[!h]
\begin{center}
%\caption{\label{example3.size}\small{The minimum model size $\cal S$ in Example 3. The quintuplet
%in each table cell consists of the 5\%, 25\%, 50\%, 75\% and 95\% quantiles of ${\cal S}$ out of 500 replications.}}
\caption{\label{example3.size} \small{ The 5\%, 25\%, 50\%, 75\% and 95\% quantiles of the minimum model size $\cal S$ out of 500 replications in Example 3.}}
\scalebox{0.8}{%
\begin{tabular}{rr|rrrrr|rrrrr}
  \hline
   & $\cal S$ &\multicolumn{5}{c|}{$p=2000$} & \multicolumn{5}{c}{$p=5000$}  \\ \hline
   & Model & 5\% & 25\% & 50\% & 75\% & 95\% & 5\% & 25\% & 50\% & 75\% & 95\% \\
\hline
$\sigma_{ij}=0.5^{|i-j|}$ & {\bf(3.a)}& 4.0 & 9.0 & 18.0 & 39.3 & 112.3  & 6.0 & 22.0 & 48.0 & 95.3 & 296.4 \\
           & {\bf(3.b)}& 6.0 & 19.0 & 43.0 & 92.0 & 253.1 & 14.0 & 45.0 & 92.5 & 198.8 & 571.6 \\
           \hline
$\sigma_{ij}=0.8^{|i-j|}$ & {\bf(3.a)}& 2.0 & 3.0 & 6.0 & 12.0 & 40.0 &  2.0 & 6.0 & 14.0 & 32.0 & 98.0 \\
           & {\bf(3.b)}& 4.0 & 4.0 & 4.0 & 6.0 & 10.0  &  4.0 & 4.0 & 5.0 & 8.0 & 18.1\\
           \hline
\end{tabular}}
\end{center}
\end{table}

\begin{table}[ht]
\begin{center}
\caption{\label{example3.prob} \small{The proportions of  ${\cal P}_{s}$ and  ${\cal P}_{a}$ in Example 3. The user-specified model sizes
$d_1 = [n/\log n]$, $d_2 = 2[n/\log n]$ and $d_3 = 3[n/\log n]$.}}
\scalebox{0.75}{%
\begin{tabular}{rr|rr|r|rrrr|r|rr|r|rrrr|r}
   \hline
&  &\multicolumn{8}{c|}{$p=2000$} & \multicolumn{8}{c}{$p=5000$}  \\
   \hline
&  &\multicolumn{3}{c|}{{\bf(3.a)}} & \multicolumn{5}{c|}{{\bf(3.b)}} & \multicolumn{3}{c|}{{\bf(3.a)}} & \multicolumn{5}{c}{{\bf(3.b)}} \\ \hline
&  &\multicolumn{2}{c|}{${\cal P}_{s}$} & ${\cal P}_{a}$ & \multicolumn{4}{c|}{${\cal P}_{s}$} & ${\cal P}_{a}$
   &\multicolumn{2}{c|}{${\cal P}_{s}$} & ${\cal P}_{a}$ & \multicolumn{4}{c|}{${\cal P}_{s}$} & ${\cal P}_{a}$ \\
   \hline
& size & $X_1$ & $X_2$ & ALL  & $X_1$ & $X_2$ & $X_{3}$ &  $X_{4}$ & ALL
       & $X_1$ & $X_2$ & ALL  & $X_1$ & $X_2$ & $X_{3}$ &  $X_{4}$ & ALL \\
   \hline
           & $d_1$ & 0.95 & 0.76 & 0.74 & 0.71 & 0.98 & 0.98 & 0.72 & 0.47   & 0.79 & 0.49 & 0.42 & 0.48 & 0.91 & 0.90 & 0.53 & 0.20  \\
$\sigma_{ij}=0.5^{|i-j|}$ & $d_2$ & 0.98 & 0.90 & 0.90 & 0.85 & 0.99 & 0.99 & 0.85 & 0.71   & 0.93 & 0.70 & 0.67 & 0.67 & 0.97 & 0.97 & 0.71 & 0.45  \\
           & $d_3$ & 1.00 & 0.95 & 0.95 & 0.91 & 0.99 & 1.00 & 0.90 & 0.81   & 0.97 & 0.81 & 0.80 & 0.75 & 0.98 & 0.99 & 0.78 & 0.55  \\
                   \hline
           & $d_1$ & 0.98 & 0.95 & 0.94 & 1.00 & 1.00 & 1.00 & 1.00 & 1.00  & 0.92 & 0.84 & 0.81 & 1.00 & 1.00 & 1.00 & 0.99 & 0.99 \\
$\sigma_{ij}=0.8^{|i-j|}$ & $d_2$ & 1.00 & 0.98 & 0.99 & 1.00 & 1.00 & 1.00 & 1.00 & 1.00  & 0.98 & 0.95 & 0.93 & 1.00 & 1.00 & 1.00 & 1.00 & 1.00 \\
           & $d_3$ & 1.00 & 1.00 & 1.00 & 1.00 & 1.00 & 1.00 & 1.00 & 1.00  & 0.99 & 0.96 & 0.96 & 1.00 & 1.00 & 1.00 & 1.00 & 1.00 \\
   \hline
\end{tabular}}
\end{center}
\end{table}

\noindent{\textbf{\bf Example 4.}}
The  Cardiomyopathy microarray dataset was once analyzed by
 \cite{Segal:Dahlquist:Conklin:2003}   and \cite{Hall:Miller:2009}. The goal
 is to identify the most influential genes for overexpression of a G
 protein-coupled receptor (Ro1) in mice.
 The response $Y$ is the Ro1 expression level, and the predictors $X_k$'s are other gene expression levels.
  Compared with the sample size $n=30$ in this dataset, the  dimension $p=6319$ is very large.

The  DC-SIS  procedure ranks two genes, labeled Msa.2134.0 and Msa.2877.0, at the top.
The scatter plots of $Y$ versus these two gene expression levels with cubic spline fit curves
in Figure 1 indicate clearly the existence of  nonlinear patterns.
Yet, our finding is  different from \cite{Hall:Miller:2009}
  in that they ranked Msa.2877.0 and Msa.1166.0 at the top  with their proposed
 generalized correlation ranking.  A natural question arises: which screening
 procedure performs better in terms of ranking? To compare the performance
 of these two procedures,
 we fit an additive model as follows:
\begin{eqnarray*}
 Y = \ell_{k1}(X_{k1}) + \ell_{k2}(X_{k2})  + \varepsilon_{k}, \textrm{ for } k = 1, 2.
\end{eqnarray*}
The DC-SIS, corresponding to $k=1$,
 regards  Msa.2134.0 and Msa.2877.0 as the two predictors, while
 the  generalized correlation ranking proposed by \cite{Hall:Miller:2009},
  corresponding to $k=2$, regards Msa.2877.0 and Msa.1166.0
 as predictors in the above model. We fit the unknown link functions $\ell_{ki}$
 using the R  \verb"mgcv" package.
The DC-SIS method clearly achieves better performance with the adjusted $R^2$ of 96.8\%
and the deviance explained of 98.3\%, in contrast to the adjusted $R^2$ of 84.5\% and
the deviance explained of 86.6\% for the generalized correlation ranking method.
{We remark here that deviance explained means the proportion of the null deviance explained by the proposed model,
 with a larger value indicating better performance.} Because both  the
 adjusted $R^2$ values and the explained deviance  are very large, it
 seems unnecessary to extract any additional  genes.

\begin{figure}[htb!]
\begin{center}
\begin{minipage}[c]{1\textwidth}
\setlength{\epsfxsize}{10.5cm} \setlength{\epsfysize}{6.0cm}
%\epsffile{Ro1Plot.eps}
\centering
\includegraphics[scale=0.6]{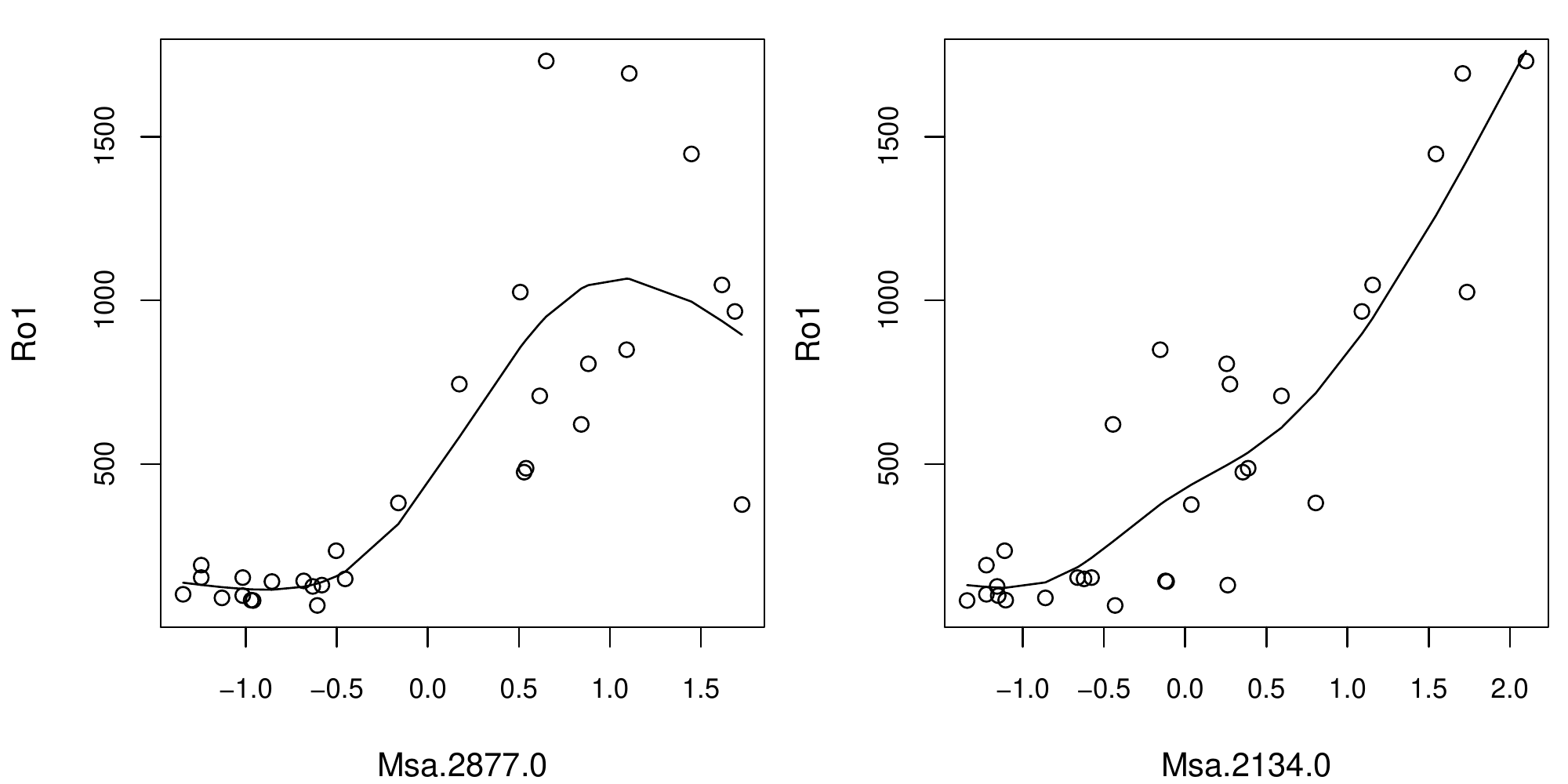}
\vskip.1cm
 {\it Figure 1. The scatter plot of $Y$ versus two genes expression levels identified
 by the DC-SIS.}
\end{minipage}
\end{center}
\end{figure}

\csection{DISCUSSION}
In this paper we proposed a sure independence screening procedure using distance correlation.
We established {the sure screening property} for this procedure
when the number of predictors diverges with  an exponential rate of the sample size.
We examined the finite-sample performance of the proposed procedure via
Monte Carlo studies and illustrated the proposed methodology through a real data example.
We followed \cite{Fan:Lv:2008} to set the cutoff $d$ in this paper and examine the effect of different
values of $d$. As pointed out by a referee, the choice of $d$ is very important  in the screening stage.
\cite{Zhao:Li:2012} proposed an approach to selecting $d$ for Cox models based on
controlling false positive rate. Their approach is merely for model-based feature screening methods.
\cite{Zhu:Li:Li:Zhu:2011} proposed an alternative method to determine $d$ for the SIRS. One
may adopt their procedure for the DC-SIS. We opt not to pursue this further. Certainly,
the selection of $d$ is similar to selection of the tuning parameter in regularization methods, and plays
an important role in practical implementation. This is a good topic for future research.

Similar to the SIS, the DC-SIS may fail to identify some important predictors which
are marginally independent of the response. Thus, it is of interest to develop an iterative
procedure to fix such an issue. In the earlier version of this paper, we proposed an iterative
version of DC-SIS. Our empirical studies including Monte Carlo simulation and real data
analysis imply that the proposed iterative DC-SIS may be used to fix the problem in a similar
spirit of ISIS \citep{Fan:Lv:2008}. Theoretical analysis of the iterative DC-SIS needs further
study. New methods to deal with identification of important predictors which are marginally
independent of the response is an important topic for future research.

\scsection{APPENDIX}
\renewcommand{\theequation}{A.\arabic{equation}}
\setcounter{equation}{0}
\scsubsection{Appendix A: Some Lemmas}

Lemmas~\ref{lem2} and \ref{lem3} will be used repeatedly in the {proof of Theorem \ref{thm:main1}}.
These two lemmas provide us two exponential inequalities, and are extracted from Lemma 5.6.1.A
and Theorem 5.6.1.A of \citet[page 200-201]{Serfling:1980}.

{\lemm\label{lem2}
Let $\mu = E(Y)$. If $\Pr\left(a\le Y\le b\right) = 1$, then
\begin{eqnarray*}
E\left[\exp\left\{s(Y-\mu)\right\}\right] \le \exp\left\{s^2(b-a)^2/8\right\}, \textrm{ for any } s>0.
\end{eqnarray*}}
%Lemma \ref{lem2} paves the road for proving Lemma \ref{lem3}.
{\lemm\label{lem3}
Let $h(Y_1,\cdots,Y_m)$ be a kernel of the $U$-statistic $U_n$, and $\theta = E\left\{h(Y_1,\cdots,Y_m)\right\}$.
If $a\le h(Y_1,\cdots,Y_m) \le b$, then, for any $t>0$ and $n\ge m$,
\begin{eqnarray*}
\pr\left(U_n-\theta\ge t\right) \le \exp\left\{-2[n/m]t^2/(b-a)^2\right\},
\end{eqnarray*}
where $[n/m]$ denotes the  integer part of $n/m$.}

Due to the symmetry of $U$-statistic, Lemma \ref{lem3} entails that
\begin{eqnarray*}
\pr\left(|U_n-\theta|\ge t\right) \le 2\exp\left\{-2[n/m]t^2/(b-a)^2\right\}.
\end{eqnarray*}

Let us introduce some notations before giving the {proof of Theorem} \ref{thm:main1}.
Let $\{\tilde{X}_k, \tilde{\y}\}$ be an independent copy of $\{X_k,\y\}$, and
define $S_{k1}= E\|X_{k}-\tilde{X}_{k}\|_1\|\y-\tilde{\y}\|_q$,
$S_{k2}= E\|X_{k}-\tilde{X}_{k}\|_1E\|\y-\tilde{\y}\|_q$, and
$S_{k3}= E\{E(\|X_{k}-\tilde{X}_{k}\|_1|X_k)E(\|\y-\tilde{\y}\|_q|\y)\}$,
and their sample counterparts
\begin{eqnarray*}
&&\widehat{S}_{k1}=\frac 1{n^2}\sum_{i,j=1}^n\|X_{ik}-X_{jk}\|_1\|\y_i-\y_j\|_q,\\
&&\widehat{S}_{k2}=\frac 1{n^2}\sum_{i,j=1}^n\|X_{ik}-X_{jk}\|_1\frac 1{n^2}\sum_{i,j=1}^n\|\y_i-\y_j\|_q,\\
&&\widehat{S}_{k3}=\frac 1{n^3}\sum_{i,j,l=1}^n\|X_{ik}-X_{lk}\|_1\|\y_j-\y_l\|_q.
\end{eqnarray*}
By definitions of distance covariance and sample distance covariance, it follows that
\[
\dcov^2(X_k,\y) = S_{k1}+S_{k2}-2S_{k3}\quad\mbox{and}\quad
\widehat{\dcov}^2(X_k,\y) = \widehat{S}_{k1}+\widehat{S}_{k2}-2\widehat{S}_{k3}.
\]

\renewcommand{\theequation}{B.\arabic{equation}}
\scsubsection{Appendix B: Proof of Theorem \ref{thm:main1}}

We aim to show the uniform consistency of the denominator and the
numerator of $\widehat{\omega}_k$ under regularity conditions respectively.
Because the denominator of $\widehat{\omega}_k$ has a similar form as the numerator,
we deal with its numerator only below. Throughout proof, the notations  $C$ and $c$
are generic constants which may take different values at
  each appearance.

%%%%%%%%%%%%%%%%%%%%%%%%%%%%%%
%%%%%%%%%%   S_k1  %%%%%%%%%%%
%%%%%%%%%%%%%%%%%%%%%%%%%%%%%%
We first deal with $\widehat S_{k1}$. Define $\widehat{S}_{k1}^\ast=\{n(n-1)\}^{-1}\underset{i\ne j}{\sum}\|X_{ik}-X_{jk}\|_1\|\y_i-\y_j\|_q$,
which is a usual $U$-statistic.
We shall establish the uniform consistency of $\widehat{S}_{k1}^\ast$ by using the theory of $U$-statistics \citep[Section 5]{Serfling:1980}. By using the Cauchy-Schwartz inequality,
\begin{eqnarray}\nonumber
S_{k1} &=& E\left(\|X_{ik}-X_{jk}\|_1\|\y_i-\y_j\|_q\right)
\le  \left\{
E\left(\|X_{ik}-X_{jk}\|_1^2\right)E\left(\|\y_i-\y_j\|_q^2\right)\right\}^{1/2}
\\\nonumber&\le&  4\left\{E(X_k^2) E\|\y\|_q^2\right\}^{1/2}.
%\label{s1bound}
\end{eqnarray}
This together with condition (C1)  implies that $S_{k1}$ is uniformly bounded in $p$, that is,
$\underset{p}{\sup}\underset{1\le k\le p}\max S_{k1}<\infty$. For any given $\varepsilon>0$,
take $n$ large enough such that $S_{k1}/n<\varepsilon$.
Then it can be easily shown that
\begin{eqnarray}\label{eqn0S1}
\begin{split}
 \pr\big(\big|\widehat S_{k1} - S_{k1}\big|\ge 2\varepsilon\big)
&=\pr\big\{\big|\widehat S_{k1}^\ast(n-1)/n - S_{k1}(n-1)/n - S_{k1}/n\big|\ge 2\varepsilon\big\}
\\&\le
\pr\big\{\big|\widehat S_{k1}^\ast - S_{k1}\big|(n-1)/n\ge 2\varepsilon - S_{k1}/n\big\}
\\
&\le
\pr\big(\big|\widehat S_{k1}^\ast - S_{k1}\big|\ge \varepsilon \big).
\end{split}
\end{eqnarray}
To establish the uniform consistency of $\widehat{S}_{k1}$, it  thus suffices to show
the uniform consistency of $\widehat S^\ast_{k1}$.
Let $h_1(X_{ik},\y_i;X_{jk},\y_j)=\|X_{ik}-X_{jk}\|_1\|\y_i-\y_j\|_q$ be the kernel of
the $U$-statistic $\widehat {S}_{k1}^\ast$. We decompose the kernel function
$h_1$ into two parts:
$h_1 = h_1{\bf1}(h_1 >M) + h_1{\bf1}(h_1 \le M)$ where $M$ will be specified later.
The $U$-statistic can now be written as follows,
\beqrs
\widehat S_{k1}^\ast %&=& \{n(n-1)\}^{-1}\sum_{i\ne j}h_1(X_{ik},\y_i;X_{jk},\y_j)\\
&=& \{n(n-1)\}^{-1}\sum_{i\ne j}h_1(X_{ik},\y_i;X_{jk},\y_j){\bf1}\left\{
h_1(X_{ik},\y_i;X_{jk},\y_j)\le M\right\}
\\&+&
\{n(n-1)\}^{-1}\sum_{i\ne j}h_1(X_{ik},\y_i;X_{jk},\y_j){\bf1}\left\{
h_1(X_{ik},\y_i;X_{jk},\y_j)> M\right\}
\\& =& \widehat S_{k1,1}^\ast + \widehat S_{k1,2}^\ast.
\eeqrs
Accordingly, we decompose $S_{k1}$ into two parts:
\begin{eqnarray*}
S_{k1} &=& E\left[h_1(X_{ik},\y_i;X_{jk},\y_j){\bf1}\left\{
h_1(X_{ik},\y_i;X_{jk},\y_j)\le M\right\}\right]
\\&+& E\left[h_1(X_{ik},\y_i;X_{jk},\y_j){\bf1}\left\{
h_1(X_{ik},\y_i;X_{jk},\y_j)> M\right\}\right]
\\&=& S_{k1,1} + S_{k1,2}.
\end{eqnarray*}
Clearly,
$\widehat S_{k1,1}^\ast$ and $\widehat S_{k1,2}^\ast$
are  unbiased estimators of $S_{k1,1}$ and $S_{k1,2}$, respectively.

We deal with the consistency of $\widehat S_{k1,1}^\ast$ first.
With the Markov's inequality, for any $t>0$, we can obtain  that
\begin{eqnarray}\nonumber%\label{eqn1}
\pr(\widehat S_{k1,1}^\ast- S_{k1,1}\ge \varepsilon)\le \exp\left(-t\varepsilon\right) \exp(-tS_{k1,1})E\{\exp(t\widehat S_{k1,1}^\ast)\}.
\end{eqnarray}
\citet[Section 5.1.6]{Serfling:1980} showed that any $U$-statistic can be represented as
  an average of averages of independent and identically distributed (i.i.d) random variables.  That is,
$\widehat S_{k1,1}^\ast = \left(n!\right)^{-1}\underset{n!}{\sum} \Omega_1(X_{1k},\y_1;\cdots;X_{nk},\y_n)$, where
 $\underset{n!}{\sum}$ denotes the summation over all possible permutations of $(1,\ldots,n)$,
and each $\Omega_1(X_{1k},\y_1;\cdots;X_{nk},\y_n)$ is an average of $m=[n/2]$ i.i.d random variables
(i.e., $\Omega_1=m^{-1}\underset{r}{\sum} h_1^{(r)}{\bf1}\{h_1^{(r)}\le M\}$).
Since the exponential function is convex, it follows from Jensen's inequality that, for $0<t\le 2s_0$,
\begin{eqnarray}\nonumber
E\{\exp(t\widehat S_{k1,1}^\ast)\}  &=&  E\big[\exp\big\{t(n!)^{-1}\sum_{n!}
\Omega_1(X_{1k},\y_1;\cdots;X_{nk},\y_n)\big\}\big] \\  \nonumber
&\le& (n!)^{-1}\sum_{n!}E\left[\exp\left\{t \Omega_1(X_{1k},\y_1;\cdots;X_{nk},\y_n)\right\}\right]
%\\\nonumber&=& E\Big\{\exp\Big(m^{-1}t\sum_r h_1^{(r)}{\bf1}\{h_1^{(r)}\le M\} \Big)\Big\}
\\\nonumber&=& E^m\big\{\exp\big(m^{-1}t h_1^{(r)}{\bf1}\{h_1^{(r)}\le M\} \big)\big\},
\end{eqnarray}
which together with Lemma \ref{lem2} entails immediately that
\begin{eqnarray}\nonumber
\pr(\widehat S_{k1,1}^\ast- S_{k1,1}\ge \varepsilon)&\le& \exp\left(-t\varepsilon\right)
E^m\big\{\exp\big(m^{-1}t  \big[h_1^{(r)}{\bf1}\{h_1^{(r)}\le M\}-S_{k1,1}\big]\big) \big\}\\\nonumber
&\le& \exp\left\{-t\varepsilon + M^2t^2/(8m) \right\}.
%\label{eqn3}
\end{eqnarray}
By choosing $t=4\varepsilon m/M^2$, we have $\pr(\widehat S_{k1,1}^\ast- S_{k1,1}\ge \varepsilon)
\leq \exp\left(-2\varepsilon^2 m /M^2 \right)$. Therefore, by the symmetry of $U$-statistic,
we can obtain easily that
\begin{eqnarray}\label{eqn5}
\pr\big(\big|\widehat S_{k1,1}^\ast- S_{k1,1}\big|\ge \varepsilon\big)
\leq 2\exp\left(-2\varepsilon^2 m /M^2 \right).
\end{eqnarray}
Next we show the consistency of $\widehat S_{k1,2}^\ast$. With Cauchy-Schwartz and Markov's inequality,
\beqrs
S_{k1,2}^2 &\le&  E\left\{h_1^2(X_{ik},\y_i;X_{jk},\y_j)\right\}\pr\left\{
h_1(X_{ik},\y_i;X_{jk},\y_j)> M\right\}
\\&\le&
E\left\{h_1^2(X_{ik},\y_i;X_{jk},\y_j)\right\}E\left[\exp\left\{
s'h_1(X_{ik},\y_i;X_{jk},\y_j)\right\}\right]/\exp\left(s'M\right),
\eeqrs
for any $s'>0$. Using the fact
$(a^2+b^2)/2 \ge  (a+b)^2/4\ge |ab|$, we have
\begin{eqnarray*}
&&
h_1(X_{ik},\y_i;X_{jk},\y_j)  =   \left\{(X_{ik} - X_{jk})^2(\y_i - \y_j)\trans (\y_i - \y_j)\right\}^{1/2}\\
&\le & 2\left\{\left(X_{ik}^2 + X_{jk}^2\right)\left(\|\y_i\|_q^2 + \|\y_j\|_q^2\right)\right\}^{1/2}
\le    \left\{\left(X_{ik}^2 + X_{jk}^2 + \|\y_i\|_q^2 + \|\y_j\|_q^2\right)^2\right\}^{1/2}\\
&  = &  X_{ik}^2 + X_{jk}^2 + \|\y_i\|_q^2 + \|\y_j\|_q^2,
\end{eqnarray*}
which yields that
\begin{eqnarray*}
E\left[\exp\left\{s'h_1(X_{ik},\y_i;X_{jk},\y_j)\right\}\right]
&\le& E\left[\exp\left\{s'\left(X_{ik}^2 + X_{jk}^2 + \|\y_i\|_q^2 + \|\y_j\|_q^2\right)\right\}\right]
%\\ &=&  E^2\left[\exp\left\{s'\left(X_{ik}^2+ \|\y_i\|_q^2\right)\right\}\right]
\\&\le& E\left\{\exp\left(2s' X_{ik}^2\right)\right\}E\left\{\exp\left(2s' \|\y_i\|_q^2\right)\right\}.
\end{eqnarray*}
The last inequality follows from the Cauchy-Schwartz inequality.
If we choose $M = cn^\gamma$ for $0<\gamma<1/2-\kappa$, then $S_{k1,2}\le \varepsilon/2$
when $n$ is sufficiently large. Consequently,
\beqr\label{eq:new1}
\pr\big(\big|\widehat S_{k1,2}^\ast - S_{k1,2}\big| >\varepsilon\big) &\le&
\pr\big(\big|\widehat S_{k1,2}^\ast\big| >\varepsilon/2\big).
\eeqr
It remains to bound the probability $\pr\big(\big|\widehat S_{k1,2}^\ast\big| >\varepsilon/2\big)$.
 We observe that the events satisfy
\beqr\label{eq:new0}
\big\{\big|\widehat S_{k1,2}^\ast\big| >\varepsilon/2\big\}
\subseteq
\big\{X_{ik}^2 + \|\y_i\|_q^2 > M/2, \textrm{ for some }  1\le i\le p \big\}.
\eeqr
To see this, we assume
that
$X_{ik}^2 + \|\y_i\|_q^2 \le M/2$ for all $1\le i\le p$. This assumption will lead to a contradiction. To be precise, under this assumption,
$h_1(X_{ik},\y_i;X_{jk},\y_j)   \le  X_{ik}^2 + X_{jk}^2 + \|\y_i\|_q^2 + \|\y_j\|_q^2\le
M$. Consequently, $\big|\widehat S_{k1,2}^\ast\big| =0$, which is a contrary to
the event
$\big|\widehat S_{k1,2}^\ast\big| >\varepsilon/2$. This verifies the  relation
(\ref{eq:new0}) is true.

%We observe that
%\beqrs
%\widehat S_{k1,2}^\ast &=& \{n(n-1)\}^{-1}\sum_{i\ne j}h_1(X_{ik},\y_i;X_{jk},\y_j){\bf1}\left\{
%h_1(X_{ik},\y_i;X_{jk},\y_j)> M\right\}\\
%&\le&  \frac{1}{n(n-1)}\sum_{i\ne j}\left(X_{ik}^2 + X_{jk}^2 + \|\y_i\|_q^2 + \|\y_j\|_q^2\right)
%{\bf1}\left(X_{ik}^2 + X_{jk}^2 + \|\y_i\|_q^2 + \|\y_j\|_q^2 > M\right)\\
%&\le&  \frac{1}{n(n-1)}\sum_{i\ne j}\left(X_{ik}^2 + X_{jk}^2 + \|\y_i\|_q^2 + \|\y_j\|_q^2\right)
%\left\{{\bf1}\left(X_{ik}^2 + \|\y_i\|_q^2 > M/2\right)+{\bf1}\left(X_{jk}^2 + \|\y_j\|_q^2 > M/2\right)\right\}\\
%&\le&\frac2n\sum_{i=1}^n\left(X_{ik}^2 + \|\y_i\|_q^2\right){\bf1}\left(X_{ik}^2 + \|\y_i\|_q^2 > M/2\right)
%\\&+&\frac2n\sum_{i=1}^n\left(X_{ik}^2 + \|\y_i\|_q^2\right)
%\frac1n\sum_{i=1}^n{\bf1}\left(X_{ik}^2 + \|\y_i\|_q^2 > M/2\right)\eeqrs
%and hence
%\beqrs
%\pr\big(\big|\widehat S_{k1,2}^\ast\big| >\varepsilon/2\big)
%\le
%n\max_{1\le k \le p}\pr\big(X_{k}^2 + \|\y\|_q^2 > M/2 \big),
%\eeqrs
By invoking condition (C1),  there must exist a constant  $C$ such that
\begin{eqnarray} \nonumber
 \pr(\|X_k\|_1^2+\|\y\|_q^2 \geq M/2)
\leq  \pr(\|X_k\|_1\geq \sqrt{M}/2) + \pr(\|\y\|_q\geq \sqrt{M}/2)
\leq  2C\exp(-s M/4).
\end{eqnarray}
The last inequality  follows from  Markov's inequality for $s>0$.
Consequently,
\beqr\nonumber
\max_{1\le k\le p}\pr\big(\big|\widehat S_{k1,2}^\ast\big| >\varepsilon/2\big)
&\le& n \max_{1\le k\le p}\pr(\|X_k\|_1^2+\|\y\|_q^2 \geq M/2)
\\\label{eq:new2}
&\le&2nC\exp(-s M/4).
\eeqr
Recall that $M = cn^{\gamma}$.
Combining the results  (\ref{eqn5}), (\ref{eq:new1}) and
 (\ref{eq:new2}), we have
\begin{eqnarray}\label{s1rate}
\pr\big(\big|\widehat S_{k1}- S_{k1}\big|\ge 4\varepsilon\big)
 \leq  2\exp\left(-\varepsilon^2 n^{1-2\gamma} \right)+2nC\exp\left(-s n^\gamma/4 \right).
\end{eqnarray}

%\newpage
%%%%%%%%%%%%%%%%%%%%%%%%%%%%%%
%%%%%%%%%%   S_k2  %%%%%%%%%%%
%%%%%%%%%%%%%%%%%%%%%%%%%%%%%%
In the sequel we turn to $\widehat S_{k2}$. We write
$\widehat S_{k2} =\widehat S_{k2,1}  \widehat  S_{k2,2}$, where
$\widehat S_{k2,1} =n^{-2}\underset{i\ne j}{\sum}\|X_{ik}-X_{jk}\|_1$, and
$\widehat S_{k2,2}=n^{-2}\underset{i\ne j}{\sum}\|\y_i-\y_j\|_q$. Similarly,
we write $S_{k2}=S_{k2,1}S_{k2,2}$, where $S_{k2,1}=E\{\|X_{ik}-X_{jk}\|_1\}$ and $S_{k2,2}=E\{\|\y_i-\y_j\|_q\}$.
%Let $\widehat S_{k2,1}^\ast =\left\{n(n-1)\right\}^{-1}\underset{i\ne j}{\operatorname{\sum}}\|X_{ik}-X_{jk}\|_1$ and
%$\widehat S_{k2,2}^\ast=\left\{n(n-1)\right\}^{-1}\underset{i\ne j}{\operatorname{\sum}}\|\y_i-\y_j\|_q$, both of which
%are second-order $U$-statistics.
Following arguments for proving (\ref{s1rate}) we can  show that
\begin{eqnarray}\label{s2rate0}
\begin{split}
& \pr\big(\big|\widehat S_{k2,1}- S_{k2,1}\big|\ge 4\varepsilon\big) \leq
 2\exp\left(-\varepsilon^2 n^{1-2\gamma} \right)+2nC\exp\left(-s n^{2\gamma}/4 \right), \textrm{ and }\\
& \pr\big(\big|\widehat S_{k2,2} - S_{k2,2}\big|\ge 4\varepsilon\big) \leq
 2\exp\left(-\varepsilon^2 n^{1-2\gamma} \right)+2nC\exp\left(-s n^{2\gamma}/4 \right).
\end{split}
\end{eqnarray}
Condition (C1) ensures that
$S_{k2,1} \leq \left\{E(\|X_{ik}-X_{jk}\|_1^2)\right\}^{1/2}
           \leq \left\{4E(X_k^2)\right\}^{1/2}$ and
$S_{k2,2} \leq \left\{E(\|\y_{i}-\y_{j}\|_q^2)\right\}^{1/2}
           \leq \left\{4E(\|\y\|_q^2)\right\}^{1/2}$  are uniformly bounded.
That is,
\[\max\big\{\max_{1\le k\le p} S_{k2,1}, S_{k2,2}\big\}\le C,\]
for some constant $C$.
Using   (\ref{s2rate0}) repetitively, we can easily prove that

\begin{eqnarray}\label{s2rate1}
\begin{split}
\pr\big\{\big|(\widehat S_{k2,1}- S_{k2,1})S_{k2,2}\big|\ge \varepsilon\big\}
&\leq \pr\big(\big|\widehat S_{k2,1}- S_{k2,1}\big|\ge \varepsilon/C\big)\\
&\leq  2\exp\left\{-\varepsilon^2 n^{1-2\gamma}/(16C^2) \right\}+2nC\exp\left(-s n^{2\gamma}/4 \right), \\
\pr\big(\big|S_{k2,1}(\widehat S_{k2,2}- S_{k2,2})\big|\ge \varepsilon \big)
&\leq \pr\big(\big|\widehat S_{k2,2}- S_{k2,2}\big|\ge \varepsilon/C\big) \\
&\leq 2\exp\left\{-\varepsilon^2 n^{1-2\gamma}/(16C^2) \right\}+2nC\exp\left(-s n^{2\gamma}/4 \right),
\end{split}
\end{eqnarray}
and
\begin{eqnarray}\label{s2rate3}
\begin{split}
&   \pr\big\{\big|(\widehat S_{k2,1} - S_{k2,1})(\widehat S_{k2,2} -S_{k2,2})\big| \ge \varepsilon\big\}
\\
\le&\pr\big(\big|\widehat S_{k2,1} - S_{k2,1}\big| \ge \sqrt{\varepsilon}\big)
+    \pr\big(\big|\widehat S_{k2,2} - S_{k2,2}\big| \ge \sqrt{\varepsilon}\big)\\
\le& 4\exp\left(-\varepsilon n^{1-2\gamma}/16 \right)+4nC\exp\left(-s n^{2\gamma}/4 \right).
\end{split}
\end{eqnarray}
It follows from Bonferroni's inequality, inequalities (\ref{s2rate1})  and (\ref{s2rate3}) that,
\begin{eqnarray}\label{s2rate}
\begin{split}
&\pr\left(\left|\widehat S_{k2} - S_{k2}\right| \ge 3\varepsilon\right)
= \pr\left(\left|\widehat S_{k2,1}\widehat S_{k2,2}- S_{k2,1}S_{k2,2}\right| \ge 3\varepsilon\right)
\\ \le&\pr\left\{\left|(\widehat S_{k2,1}- S_{k2,1})S_{k2,2}\right| \ge \varepsilon\right\}
     +\pr\left\{\left|S_{k2,1}(\widehat S_{k2,2}- S_{k2,2})\right| \ge \varepsilon\right\}
\\  & +\pr\left\{\left|(\widehat S_{k2,1}- S_{k2,1})(\widehat S_{k2,2}-S_{k2,2})\right| \ge \varepsilon\right\}
%\\  \le& 4\exp\left\{-\varepsilon^2 n^{1-2\gamma}/(16C^2) \right\}+4nC\exp\left(-s n^{2\gamma}/4 \right)
%%\\   &   +2\exp\left\{-\varepsilon^2 n^{1-2\gamma}/(16C^2) \right\}+2nC\exp\left(-s n^{2\gamma}/4 \right)
%\\   &   +4\exp\left(-\varepsilon n^{1-2\gamma}/16 \right)+4nC\exp\left(-s n^{2\gamma}/4 \right)
\\  \le& 8\exp\left\{-\varepsilon^2 n^{1-2\gamma}/(16C^2) \right\}+8nC\exp\left(-s n^{2\gamma}/4 \right),
\end{split}
\end{eqnarray}
where the last inequality holds when  $\varepsilon$ is sufficiently small and $C$ is sufficiently large.

%%%%%%%%%%%%%%%%%%%%%%%%%%%%%%%%
%%%%%%%%%%%  S_k3  %%%%%%%%%%%%%
%%%%%%%%%%%%%%%%%%%%%%%%%%%%%%%%
It remains to the uniform consistency of  $\widehat S_{k3}$. We first study the following  $U$-statistic:
\begin{eqnarray}\nonumber
\widehat{S}_{k3}^\ast&=&\frac{1}{n(n-1)(n-2)}\sum_{i<j<l}\Big\{
\|X_{ik}-X_{jk}\|_1\|\y_j-\y_l\|_q + \|X_{ik}-X_{lk}\|_1\|\y_j-\y_l\|_q+\\\nonumber&&
\hspace{4.18cm}\|X_{ik}-X_{jk}\|_1\|\y_i-\y_l\|_{q}
+\|X_{lk}-X_{jk}\|_1\|\y_i-\y_l\|_q+\\\nonumber&&
\hspace{4.18cm}
\|X_{lk}-X_{jk}\|_1\|\y_i-\y_j\|_q
+\|X_{lk}-X_{ik}\|_1\|\y_i-\y_j\|_q\Big\}\\
&=:& \frac{6}{n(n-1)(n-2)}\sum_{i<j<l} h_3(X_{ik},\y_i;X_{jk},\y_j;X_{lk},\y_l).
\end{eqnarray}
Here, $h_3(X_{ik},\y_i;X_{jk},\y_j;X_{lk},\y_l)$ is the kernel of $U$-statistic $\widehat S_{k3}^\ast$.
Following the arguments to deal with $\widehat{S}_{k1}^\ast$, we decompose $h_3$
into two parts: $h_3=h_3{\bf1}(h_3 >M) + h_3{\bf1}(h_3 \le M)$.
Accordingly,
\begin{eqnarray*}
\widehat{S}_{k3}^\ast &=& \frac{6}{n(n-1)(n-2)}\sum_{i<j<l} h_3{\bf1}(h_3\le M)+ \frac{6}{n(n-1)(n-2)}\sum_{i<j<l} h_3{\bf1}(h_3 >M)\\
&=& \widehat S_{k3,1}^\ast + \widehat S_{k3,2}^\ast,\\
{S}_{k3} &=& E\left\{h_3{\bf1}(h_3\le M)\right\}+ E\left\{ h_3{\bf1}(h_3 >M)\right\} = S_{k3,1} + S_{k3,2}.
\end{eqnarray*}
Following  similar arguments for proving (\ref{eqn5}), we can show that
\begin{eqnarray}\label{sk31rate}
\pr\big(\big|\widehat S_{k3,1}^\ast- S_{k3,1}\big|\ge \varepsilon\big)
\leq 2\exp\left(-2\varepsilon^2 m' /M^2 \right),
\end{eqnarray}
where $m'= [n/3]$ because $\widehat S_{k3,1}^\ast$ is a third-order $U$-statistic.

Then we deal with $\widehat S_{k3,2}^\ast$. We observe  that $h_3(X_{ik},\y_i;X_{jk},\y_j;X_{lk},\y_l) \le 4(X_{ik}^2+X_{jk}^2+X_{lk}^2+\|\y_i\|_q^2 + \|\y_j\|_q^2 + \|\y_l\|_q^2)/6$, which will be smaller than $M$ if $X_{ik}^2 + \|\y_i\|_q^2 \le M/2$ for all $1\le i\le p$.  Thus, for any $\varepsilon>0$, the events satisfy
\beqrs
\big\{\big|\widehat S_{k3,2}^\ast\big| >\varepsilon/2\big\}
\subseteq
\big\{X_{ik}^2 + \|\y_i\|_q^2 > M/2, \textrm{ for some }  1\le i\le p \big\}.
\eeqrs
By using the similar arguments to prove (\ref{eq:new2}), it follows that
\beqr\label{sk32rate}
\pr\big(\big|\widehat S_{k3,2}^\ast - S_{k3,2}\big| >\varepsilon\big)  \le
\pr\big(\big|\widehat S_{k3,2}^\ast\big| >\varepsilon/2\big) \le  2nC\exp(-s M/4).
\eeqr
Then, we combine the results (\ref{sk31rate}) and (\ref{sk32rate}) with $M = cn^{\gamma} $ for some $0<\gamma<1/2-\kappa$
to obtain that
\begin{eqnarray}\label{s3rate1}
\pr\left(\left|\widehat S_{k3}^\ast- S_{k3}\right| \ge  2\varepsilon\right)&\leq&
2\exp\left(-2\varepsilon^2 n^{1-2\gamma}/3 \right)+2nC\exp\left(-s n^\gamma/4 \right).
\end{eqnarray}
By the definition of $\widehat S_{k3}$,
\[
\widehat S_{k3}=\frac{(n-1)(n-2)}{n^2}\left\{\widehat S_{k3}^\ast  + \frac{1}{(n-2)}\widehat S_{k1}^\ast\right\}.
\]
Thus, using similar techniques to deal with $\widehat S_{k1}$, we can obtain that
\begin{eqnarray*}
 \pr\left(\left|\widehat S_{k3}-S_{k3}\right|\ge 4\varepsilon\right)
 &=&  \pr\left\{\left|\frac{(n-1)(n-2)}{n^2}\left(\widehat S_{k3}^\ast - S_{k3}\right) -\frac{3n-2}{n^2}S_{k3}\right.\right. \\
 & &  ~~~~~~~\left.\left.+~\frac{n-1}{n^2}\left(\widehat S_{k1}^\ast- S_{k1}\right)+  \frac{n-1}{n^2}S_{k1} \right|\ge 4\varepsilon\right\}.
\end{eqnarray*}
Using similar arguments for dealing with $S_{k1}$, we can show that $S_{k3}$ is uniformly bounded in $p$.
Taking  $n$ large enough such that $\{(3n-2)/n^2\}S_{k3}\le \varepsilon$ and
$\{(n-1)/n^2\}S_{k1}\le \varepsilon$,   then
%the right hand in the above display is less than or equal to
%\begin{equation}
%\pr\big(\big|\widehat S_{k3}-S_{k3}\big|\ge 4\varepsilon\big)\le \pr\big(\big|\widehat S_{k3}^\ast - S_{k3}\big|\ge \varepsilon\big)
%+ \pr\big\{\big|\widehat S_{k1}^\ast - S_{k1}\big|\ge \varepsilon \big\}.
%\label{t1}
%\end{equation}
\begin{eqnarray} \label{s3rate}
\begin{split}
\pr\big(\big|\widehat S_{k3}-S_{k3}\big|\ge 4\varepsilon\big)
&\le \pr\big(\big|\widehat S_{k3}^\ast - S_{k3}\big|\ge \varepsilon\big)
+ \pr\big\{\big|\widehat S_{k1}^\ast - S_{k1}\big|\ge \varepsilon \big\}
\\
%&\le 2\exp\left(-\varepsilon^2 n^{1-2\gamma}/6 \right)+2nC\exp\left(-s n^\gamma/4 \right) \\
%&+2\exp\left(-\varepsilon^2 n^{1-2\gamma}/4 \right)+2nC\exp\left(-s n^\gamma/4 \right)   \\
&\le 4 \exp\left(-\varepsilon^2 n^{1-2\gamma}/6 \right)+4nC\exp\left(-s n^\gamma/4 \right).
\end{split}
\end{eqnarray}
The last inequality follows from (\ref{s1rate}) and (\ref{s3rate1}).
This, together with (\ref{s1rate}), (\ref{s2rate}) and the Bonferroni's inequality, implies
\begin{eqnarray}\label{numrate}
\begin{split}
&\pr\big\{\big|(\widehat S_{k1}+\widehat S_{k2}-2\widehat S_{k3})
- \left(S_{k1} + S_{k2} - 2S_{k3}\right)\big| \ge \varepsilon\big\}\\
\le& \pr\big(\big|\widehat S_{k1}
- S_{k1}\big| \ge \varepsilon/4\big)
+ \pr\big(\big|\widehat S_{k2}
- S_{k2}\big| \ge \varepsilon/4\big)
+ \pr\big(\big|\widehat S_{k3}
- S_{k3}\big| \ge \varepsilon/4\big)\\
%&\le&  2\exp\left[-(\varepsilon/8)^2 n^{1-2\gamma} \right]+2nC\exp\left(-s n^\gamma/4 \right) \\ \nonumber
%&& +~8\exp\left[-(\varepsilon/12)^2 n^{1-2\gamma}/(4C^2) \right]+8nC\exp\left(-s n^{2\gamma}/4 \right)  \\ \nonumber
%&& +~ 4 \exp\left[-2(\varepsilon/16)^2 n^{1-2\gamma}/3 \right]+4nC\exp\left(-s n^\gamma/4 \right) \\
=&  O\left\{ \exp\left(-c_1\varepsilon^2 n^{1-2\gamma} \right)+n\exp\left(-c_2n^\gamma \right) \right\},
\end{split}
\end{eqnarray}
 for some positive constants $c_1$ and $c_2$.
The convergence rate of the numerator of $\widehat\omega_k$ is now achieved.
Following similar arguments, we can obtain the convergence rate of the denominator.
In effect the convergence rate of $\widehat\omega_k$
has the same form of (\ref{numrate}). We omit the details here.
%Hence, for any $M>0$, there exist positive constants $c_1>0$ and $c_2>0$ such that
%\begin{eqnarray} \label{omegarate}
%\pr\left\{\left|\widehat\omega_k - \omega_k\right| \ge \varepsilon\right\}
%\leq  O\left\{ \exp\left(-c_1\varepsilon^2 n^{1-2\gamma} \right)+n\exp\left(-c_2n^\gamma \right)  \right\}.
%\end{eqnarray}
Let $\varepsilon=cn^{-\kappa}$, where $\kappa$ satisfies $0<\kappa+\gamma<1/2$. We thus have

\begin{eqnarray}\nonumber%\label{thm:pt1}
\begin{split}
\pr\big\{\max_{1\le k\le p}\left|\widehat\omega_k - \omega_k\right| \ge cn^{-\kappa}\big\}
&\le p\max_{1\le k\le p}\pr\left\{\left|\widehat\omega_k - \omega_k\right| \ge cn^{-\kappa}\right\}   \\
&\le O\left( p\left[\exp\left\{-c_1 n^{1-2(\kappa+\gamma)} \right\}+n\exp\left(-c_2 n^{\gamma}\right)\right] \right).
\end{split}
\end{eqnarray}
The first part of Theorem \ref{thm:main1} is proven.

Now we deal with the second part of Theorem \ref{thm:main1}.
If $\calA \nsubseteq \widehat{\calA}^\star$, then there must exist
some $k\in\calA$ such that
$\widehat \omega_k < cn^{-\kappa}$.
It follows
 from condition (C2)  that $|\widehat \omega_k - \omega_k| > cn^{-\kappa}$ for some $k\in\calA$, indicating that
the events satisfy
$\big\{\calA \nsubseteq \widehat{\calA}^\star\big\}
\subseteq
\big\{|\widehat \omega_k - \omega_k| > cn^{-\kappa}, \textrm{ for some }
 k\in\calA\big\}$,   and hence $\mathcal{E}_n = \big\{\underset{k\in \calA}{\max}\left|\widehat\omega_k - \omega_k\right|\le cn^{-\kappa}  \big\}
\subseteq
 \big\{\calA \subseteq \widehat{\calA}^\star\big\}.$
Consequently,
\begin{eqnarray*}\nonumber
\pr(\calA \subseteq \widehat{\calA}^\star)
&\ge& \pr(\mathcal{E}_n)=1-\pr(\mathcal{E}_n^c)
 =  1- \pr \big(\min_{k\in \calA}\left|\widehat\omega_k - \omega_k\right|\ge cn^{-\kappa}  \big)\\ \nonumber
&=& 1- s_n\pr \left\{\left|\widehat\omega_k - \omega_k\right|\ge cn^{-\kappa}  \right\}
\\&\ge&  1- O\left(s_n\left[\exp\left\{-c_1 n^{1-2(\kappa+\gamma)} \right\}+n\exp\left(-c_2 n^{\gamma}\right)\right]\right),
\end{eqnarray*}
where $s_n$ is the cardinality of $\calA$. This completes the proof of the second part.
\hfill$\fbox{}$

\scsection{REFERENCES}

%%%%%%%%%%%%%%%%%%%%%%%%%%%%%%%%%%%%%%%%%%%%%%%%%%%%%%%%%%%%%%%%%%%%%%%%%%%%%%%%%%%%%%%%%%%%%%%%%%%%%%%%%%%%%%%%%%%%%%%%%%%%

\renewcommand{\baselinestretch}{1.00}
\baselineskip=14pt
\begin{description}

\newcommand{\enquote}[1]{``#1''}
\expandafter\ifx\csname
natexlab\endcsname\relax\def\natexlab#1{#1}\fi
%\bibitem[{Akaike(1973)}]{Akaike:1973}
%Akaike, H. (1973) \enquote{Maximum likelihood identification of Gaussian
%autoregressive moving average models}
% \textit{Biometrika}
%{\bf60} 255--265.

\bibitem[{Ashburner, et al.(2000)}]{Ashburneretal:2000}
Ashburner, M., Ball, C. A., Blake, J. A., Botstein, D., Butler, H., et al. (2000),
\enquote{Gene Ontology: Tool for the Unification of Biology. The Gene Ontology
Consortium,} \textit{Nature Genetics}, {\bf 25}, 25-–29.

%\bibitem[{Bickel and Levina(2008)}]{Bickel:Levina:2008}
%Bickel, P. J.,   and Levina E. (2008), \enquote{Regularized estimation of
%large covariance matrices,}
% \textit{Annals of Statistics},  {\bf36},  199--227.

\bibitem[{Bild, et al.(2006)}]{Bildetal:2006}
Bild, A., Yao, G., Chang, J. T., Wang, Q., Potti, A., et al.
%Chasse, D., Joshi, M.-B., Harpole, D., Lancaster, J. M.,
%Berchuck, A., Olson, J. A., Marks, J. R., K., D. H., West, M. and Nevins, J. R.
(2006), \enquote{Oncogenic pathway signatures in human cancers as a guide to targeted therapies,} \textit{Nature} {\bf 439} 353-357.

%\bibitem[{Breiman(1995)}]{Breiman:1995}
%Breiman, L. (1995) \enquote{Better subset regression using the nonnegative
%garrote}. {\it Technometrics}  {\bf 37}  373--384.

\bibitem[{Candes and Tao(2007)}]{Candes:Tao:2007}
Candes, E. and Tao, T. (2007),  \enquote{The Dantzig selector: statistical
estimation when $p$ is much larger than $n$ (with discussion),}
\textit{Annals of Statistics},  {\bf 35},   2313--2404.

\bibitem[{Chen, et al.(2011)}]{Chenetal:2011}
Chen, L. S., Paul, D., Prentice, R. L. and Wang, P. (2011), \enquote{A regularized Hotelling's $T^2$ test for pathway analysis
in proteomic studies,} \textit{Journal of the American Statistical Association}  {\bf 106} 1345--1360.

%\bibitem[{Chiang, et al.(2006)}]{Chiang:2006}
%Chiang, A. P., Beck, J. S., Yen, H.-J., Tayeh, M. K., Scheetz, T. E., et al. (2006),
%Swiderski, R., Nishimura, D., Braun, T. A., Kim, K.-Y., Huang, J., Elbedour, K., Carmi, R., Slusarski, D. C.,
%Casavant, T. L., Stone, E. M. and Shefield, V. C. (2006),
%\enquote{Homozygosity Mapping with
%SNP Arrays Identifies a Novel Gene for Bardet-Biedl Syndrome (BBS10),}
%\textit{Proceeding of the National Academy of Sciences}, {\bf103}, 6287--6292.

%\bibitem[{Cook and Forzani(2009)}]{Cook:Forzani:2009}
%Cook, R. D. and  Forzani, L. (2009)
%\enquote{Likelihood-based sufficient dimension reduction}
%\textit{Journal of the American Statistical Association}  {\bf 104} 197--208.

%\bibitem[(Dinu, et al. (2007)}]{Dinuetal:2011}
%Dinu, I., Potter, J. D., Mueller, T., Liu, Q., Adewale, A. J., et al. (2007), Improving
%Gene Set Analysis of Microarray Data by SAM-GS, \textit{BMC Bioinformatics},
%{\bf 8}, 242.

\bibitem[{Efron, Hastie, Johnstone and Tibshirani(2004)}]{Efron:Hastie:Johnstone:Tibshirani:2004}
Efron, B., Hastie T., Johnstone, I. and Tibshirani, R. (2004),
\enquote{Least angle regression (with discussion),}
\textit{Annals of Statistics},  {\bf 32}, 409--499.

\bibitem[{Efron and Tibsirani(2007)}]{Efron:Tibshirani:2007}
Efron, B., and Tibshirani, R. (2007), \enquote{On Testing the Significance of Sets of
Genes,} \textit{The Annals of Applied Statistics}, {\bf 1}, 107-–129.

%\bibitem[{Fan and Fan(2008)}]{Fan:Fan:2008}
%Fan, J. and Fan, Y. (2008). \enquote{High-dimensional classification using
%features annealed independence rules}  \textit{The Annals of
%Statistics}, {\bf 36} 2605--2637.

{\bibitem[{Fan, Feng and Song(2011)}]{Fan:Feng:Song:2011}
Fan, J., Feng, Y. and Song, R. (2011),  \enquote{Nonparametric independence
screening in sparse ultra-high dimensional additive models,}
\textit{Journal of the American Statistical Association},
{\bf106}, 544--557.}

%\bibitem[{Fan, Feng and Wu(2010)}]{Fan:Feng:Wu:2010}
%Fan, J., Feng, Y. and Wu, Y. (2010),  \enquote{Ultrahigh dimensional variable
%selection for Cox's proportional hazards model,} IMS Collection,
%{\bf6}, 70--86.
%\textit{This manuscript is available at
%http://www.orfe.princeton.edu/~jqfan/fan/manuscripts.html}

\bibitem[{Fan and Li(2001)}]{Fan:Li:2001}
Fan, J.,   and Li, R.   (2001),  \enquote{Variable selection via nonconcave
penalized likelihood and it oracle properties,}
 \textit{Journal of the American Statistical Association},
  {\bf96}, 1348--1360.

\bibitem[{Fan and Lv(2008)}]{Fan:Lv:2008}
Fan, J.  and Lv, J. (2008),  \enquote{Sure independence screening for
ultrahigh dimensional feature space (with discussion),}
\textit{Journal of the Royal Statistical Society, Series B},
{\bf 70}, 849--911.

%\bibitem[{Fan and Lv(2010)}]{Fan:Lv:2010}
%Fan, J.  and Lv, J. (2010)  \enquote{A selective overview of variable
%selection in high dimensional feature space}. \textit{Statistica
%Sinica}  {\bf 20}   101--148.

\bibitem[{Fan, Samworth and Wu(2009)}]{Fan:Samworth:Wu:2009}
Fan, J., Samworth, R. and Wu, Y. (2009),   \enquote{Ultrahigh dimensional
feature selection: beyond the linear model,} \textit{Journal of
Machine Learning Research}, {\bf 10}, 1829--1853.

\bibitem[{Fan and Song(2010)}]{Fan:Song:2009}
Fan, J. and Song, R. (2010),   \enquote{Sure independence screening in
generalized linear models with NP-dimensionality,}
\textit{The Annals of Statistics}, {\bf38}, 3567--3604.

%\bibitem[{Foster and George(1994)}]{Foster:George:1994}
%Foster, D. P.  and George, E. I. (1994) \enquote{The risk inflation criterion
%for multiple regression}
% \textit{Annals of Statistics}  {\bf22}  1947--1975.

%\bibitem[{Frank  and Friedman(1993)}]{Frank:Friedman:1993}
%Frank, I. E. and Friedman, J. H. (1993)  \enquote{A statistical view of some
%chemometrics regression tools}  \textit{Technometrics} {\bf35}
%109--148.

\bibitem[{Hall and Miller(2009)}]{Hall:Miller:2009}
Hall, P. and Miller, H. (2009), \enquote{Using generalized correlation to
effect variable selection in very high dimensional problems,}
\textit{Journal of Computational and Graphical Statistics},
{\bf 18}, 533--550.

%\bibitem[{Huang, Ma and Zhang(2008)}]{Huang:Ma:Zhang:2008}
%Huang, J., Ma, S. G. and Zhang, C. H. (2008),
%\enquote{Adaptive Lasso for sparse high-dimensional regression models,}
%\textit{Statistica Sinica}, \textbf{18},  1603--1618

\bibitem[{Ji and Jin(2012)}]{Ji:Jin:2010}
Ji, P. and Jin, J. (2012),
\enquote{UPS delivers optimal phase diagram in high dimensional variable selection,}
\textit{Annals of Statistics}, {\bf40}, 73-103.

\bibitem[{Jones, et al.(2008)}]{Jonesetal:2008}
Jones, S., Zhang, X., Parsons, D. W., Lin, J. C.-H., Leary, R. J., et al.
%Angenendt, P., Mankoo, P., Carter, H.,
%Kamiyama, H., Jimeno, A., Hong, S.-M., Fu, B., Lin, M.-T., Calhoun, E. S., Kamiyama, M.,
%Walter, K., Nikolskaya, T., Nikolsky, Y., Hartigan, J., Smith, D. R., Hidalgo, S. D. M. Leach, Klein, A. P.,
%Jaffee, E. M., Goggins, M., Maitra, A., Iacobuzio-Donahue, J. R. C. amd Eshleman, Kern,S.E., Hruban, R. H.,
%Karchin, R., Papadopoulos, N., Parmigiani, G., Vogelstein, B., Velculescu, V. E. and Kinzler, K. W.
(2008), \enquote{Core Signaling Pathways in Human Pancreatic Cancers Revealed by Global Genomic Analyses,} \textit{Science}, {\bf 321} 1801.

%\bibitem[{Kanehisa(2002)}]{Kanehisa:2002}
%Kanehisa, M. (2002), The KEGG Database, \textit{Novartis Foundation Symposium},
%{\bf 247}, 91-–101.

%\bibitem[{Li, Cook and Nachtsheim(2005)}]{Li:Cook:Nachtsheim:2005}
%Li, L., Cook, R.D., and Nachtsheim, C.J. (2005),
% \enquote{Model-free variable selection,} \textit{Journal of the Royal Statistical Society, Series B},
%{\bf 67},  285--299.

\bibitem[{Kim, Choi and Oh(2008)}]{Kim:Choi:Oh:2008}
Kim, Y., Choi, H. and Oh, H. S. (2008),
 \enquote{Smoothly clipped absolute deviation on high dimensions,}
 \textit{Journal of the American Statistical Association,}
{\bf 103},  1665--1673.

\bibitem[{Mootha, et al.(2003)}]{Moothaetal:2003}
Mootha, V. K., Lindgren, C. M., Eriksson, K. F., Subramanian, A., Sihag, S., et
al. (2003),  \enquote{PGC-1-Responsive Genes Involved in Oxidative Phosphorylation
Are Coordinately Downregulated in Human Diabetes,} \textit{Nature Genetics},
{\bf 34}, 267-273.

%\bibitem[{Pollard(1984)}]{Pollard:1984}
%Pollard, D. (1984) \textit{Convergence of Stochastic Processes}, Springer, New York.

%\bibitem[{Scheetz, et al.(2006)}]{Scheetz:2006}
%Scheetz, T. E., Kim, K.-Y. A., Swiderski, R. E., Philp1, A. R., Braun, T. A., et al. (2006),
%Knudtson, K. L., Dorrance, A. M., DiBona, G. F., Huang, J., Casavant, T. L., Shefield, V. C. and Stone,
%E. M. (2006),
%\enquote{Regulation of gene expression in the mammalian eye and its relevance to eye
%disease,} \textit{Proceeding of the National Academy of Sciences}, {\bf103}, 14429--14434.

%\bibitem[{Schwartz(1978)}]{Schwartz:1978}
%Schwartz, G. (1978)  \enquote{Estimating the dimension of a model}
%\textit{Annals of  Statistics} {\bf6}  461--464.

\bibitem[{Segal, Dahlquist and Conklin(2003)}]{Segal:Dahlquist:Conklin:2003}
Segal, M. R., Dahlquist, K. D., and Conklin, B. R. (2003),
\enquote{Regression approach for microarray data analysis,}
\textit{Journal of Computational Biology,} {\bf10}, 961--980.

\bibitem[{Serfling(1980)}]{Serfling:1980}
Serfling, R.~J. (1980),   {\it Approximation Theorems of
Mathematical Statistics,} New York: John Wiley \& Sons Inc.

\bibitem[{Subramanian, et al.(2005)}]{Subramanianetal:2005}
Subramanian, A., Tamayo, P., Mootha, V. K., Mukherjee, S., Ebert, B. L., et
al. (2005), \enquote{Gene Set Enrichment Analysis: A Knowledge-Based Approach
for Interpreting Genome-Wide Expression Profiles,} \textit{Proceedings of the National
Academy of Sciences of the USA}, {\bf 102}, 15545-15505.

\bibitem[{Szekely and  Rizzo(2009)}]{Szekely:Rizzo:2009}
Sz\'{e}kely, G.  J. and  Rizzo, M. L.
(2009), \enquote{Brownian distance covariance,}
 \textit{Annals of Applied Statistics}, {\bf3}, 1233--1303.

\bibitem[{Szekely,  Rizzo and  Bakirov(2007)}]{Szekely:Rizzo:Bakirov:2007}
Sz\'{e}kely, G.  J.,  Rizzo, M. L. and  Bakirov, N. K. (2007),
\enquote{Measuring and testing dependence by correlation of distances,}
 \textit{Annals of Statistics},  {\bf35},  2769--2794.

\bibitem[{Tian, et al.(2005)}]{Tianetal:2005}
Tian, L., Greenberg, S. A., Kong, S. W., Altschuler, J., Kohane, I. S., and Park,
P. J. (2005), \enquote{Discovering Statistically Significant Pathways in Expression
Profiling Studies,} \textit{Proceedings of the National Academy of Sciences of the
USA}, {\bf 102}, 13544-13549.

\bibitem[{Tibshirani(1996)}]{Tibshirani:1996}
Tibshirani, R. (1996),  \enquote{Regression shrinkage and selection via LASSO,}
\textit{Journal of the Royal  Statistical Society, Series B}, {\bf58},
267--288.

\bibitem[{Wang(2009)}]{Wang:2009}
Wang, H. (2009), \enquote{ Forward regression for ultra-high dimensional
variable screening,}
\textit{Journal of the American Statistical Association}, {\bf104}, 1512--1524.

\bibitem[{Zhao and Li (2012)}]{Zhao:Li:2012}
Zhao, S. D. and Li, Y. (2012), \enquote{Principled sure independence screening for Cox models with ultra-high-dimensional covariates,}
\textit{Journal of Multivariate Analysis}, {\bf 105}, 397--411.

%\bibitem[{Zhao and Yu(2006)}]{Zhao:Yu:2006}
%Zhao, P. and Yu, B. (2006), \enquote{On model selection consistency
%of Lasso}, \textit{Journal of Machine Research}  {\bf7}
%  2541--2563.

{\bibitem[{Zhu, Li, Li and Zhu(2011)}]{Zhu:Li:Li:Zhu:2011}
Zhu, L. P., Li, L., Li, R. and Zhu, L. X. (2011), \enquote{Model-free feature
screening for ultrahigh dimensional data,}
\textit{Journal of the American Statistical Association}, {\bf106}, 1464--1475.}

\bibitem[{Zou(2006)}]{Zou:2006}
Zou, H.  (2006), \enquote{The adaptive lasso and its oracle properties,}
\textit{Journal of the American Statistical Association}, {\bf101}, 1418--1429.

\bibitem[{Zou and Hastie(2005)}]{Zou:Hastie:2005} Zou, H. and Hastie, T. (2005),
\enquote{Regularization and variable selection via the elastic net,}
\textit{Journal of the Royal  Statistical Society, Series B}, {\bf 67}, 301--320.

\bibitem[{Zou and Li(2008)}]{Zou:Li:2008} Zou, H. and Li, R. (2008),
\enquote{One-step sparse estimates in nonconcave penalized likelihood models,}
\textit{Annals of Statistics}, {\bf36}, 1509--1533.

\bibitem[{Zou and Zhang(2009)}]{Zou:Zhang:2009} Zou, H. and Zhang, H. H. (2009),
\enquote{On the adaptive elastic-net with a diverging number of parameters,}
\textit{Annals of Statistics}, {\bf37}, 1733-1751.

\end{description}

\end{document}